\documentclass[draft]{agujournal2019}
\usepackage{url} 
\usepackage{lineno}
\usepackage{amssymb,amsmath}
\usepackage{soul}
\draftfalse
\journalname{Space Weather}

\begin{document}

\title{The impact of non-equilibrium plasma distributions on solar wind measurements by Vigil's Plasma Analyser}

\authors{H.~Zhang\affil{1,2}, D.~Verscharen\affil{1}, and G.~Nicolaou\affil{1}}

\affiliation{1}{Mullard Space Science Laboratory, University College London, Dorking, RH5~6NT, UK}

\affiliation{2}{Deep Space Exploration Laboratory, Beijing, 100195, China}

\correspondingauthor{Daniel Verscharen}{d.verscharen@ucl.ac.uk}

\begin{keypoints}
\item ESA's Vigil mission will measure the solar wind from the fifth Lagrange point for space weather monitoring.
\item We study the impact of non-equilibrium plasma distributions on the performance of the Vigil/PLA instrument.
\item Under reasonable solar wind conditions, non-equilibrium distributions can deteriorate onboard moments. Fitted moments are still reliable.
\end{keypoints}

%


\begin{abstract}
In order to protect society from space weather impacts, we must monitor space weather and obtain early warnings for extreme events if possible. For this purpose, the European Space Agency is currently preparing to launch the Vigil mission towards the end of this decade as a space-weather monitor at the fifth Lagrange point of the Sun--Earth system. Vigil will carry, amongst other instruments, the Plasma Analyzer (PLA) to provide quasi-continuous measurements of solar wind ions. We model the performance of the PLA instrument, considering typical solar wind plasma conditions,  to compare the expected observations of PLA with the assumed input conditions of the solar wind. We evaluate the instrument performance under realistic, non-equilibrium plasma conditions, accounting for temperature anisotropies, proton beams, and the contributions from drifting $\alpha$-particles. We examine the accuracy of the instrument's performance over a range of input solar wind moments.  We identify sources of potential errors due to non-equilibrium plasma conditions and link these to instrument characteristics such as its angular and energy resolution and its field of view. We demonstrate the limitations of the instrument and potential improvements such as applying ground-based fitting techniques to obtain more accurate measurements of the solar wind even under non-equilibrium plasma conditions. The use of ground processing of plasma moments instead of on-board processing is crucial for the extraction of reliable measurements. 
\end{abstract}

\section*{Plain Language Summary}
Space weather originates at the Sun and affects human life. An effective space-weather monitor is required to detect severe space weather events and provide early warnings before such events arrive at Earth. The European Space Agency's (ESA's) Vigil mission will carry the Plasma Analyser (PLA) instrument to obtain measurements of the solar wind proton moments such as their number density, velocity, and temperature. We predict the expected performance of the PLA instrument by modelling its response to realistic solar wind conditions, which accounts for non-equilibrium effects such as temperature anisotropy, proton beams, and $\alpha$-particles. We also study the impact of other non-equilibrium distributions such as $\kappa$-distributions (in the Appendix) to quantify the performance over a wide range of expected plasma conditions. We quantify the measurement accuracy by comparing the input and output parameters of the model and discuss possible improvements to the analysis of data from Vigil/PLA.

\section{Introduction}

Space weather severely affects local and global infrastructure on Earth and in near-Earth space \cite<e.g.,>{Hapgood_2011,SCHRIJVER_2015,Nicolaou_2020}. According to previous estimates, the total financial damage associated with a severe space-weather event is of order  333.7 billion US-dollars, assuming that the event leads to a power outage with a duration of 12 months \cite<e.g.,>{Oughton_2016,Eastwood_2017,Thomas_2018}.
Activity at the solar source, the propagation of space weather events through interplanetary space, and the Earth's response to these events are the main aspects determining space weather and its impact on humanity \cite{Schwenn_2006,Cranmer_2017,Eastwood_2017b,Temmer_2021}. There are various types of solar activity that primarily drive severe space weather.  Coronal mass ejections (CMEs) and  co-rotating interaction regions (CIRs), for example, can drive significant space-weather phenomena \cite{Thomas_2018}. In addition, fast solar wind streams can cause enhancements in the Earth's radiation belts when directed towards the Earth \cite<e.g.,>{Baker_2013,LiouKan_2014,Thomas_2018}.

In response to these space weather risks, satellite-based monitoring and forecasting systems have become increasingly important. Examples for space missions with space-weather capabilities include the Solar TErrestrial RElations Observatory (STEREO) \cite<e.g.,>{dekoning_2011,Mishra_2013} and the Advanced Composition Explorer (ACE) \cite<e.g.,>{Stone_1998}. 
The European Space Agency's (ESA's) Vigil mission, formerly known as the Lagrange mission, is currently being developed with the goal to provide quasi-continuous monitoring of the solar source regions of space weather and the heliospheric environment to provide early warnings of potentially hazardous space weather. The Vigil spacecraft will orbit the Sun at the fifth Sun--Earth Lagrange point (L5), which is located at a heliocentric distance of 1\,au but constantly $60^{\circ}$ behind the Earth’s orbit. Due to the Sun's synodic rotation period in the L5 reference frame, the Vigil  spacecraft will face regions of the Sun's surface about 4 to 5 days before they point toward Earth \cite{Thomas_2018}.

Vigil is expected to carry, in addition to its remote-sensing payload, an in-situ Plasma Analyser (PLA) instrument and a fluxgate magnetometer (MAG). PLA is an electrostatic analyzer that will measure the protons of the solar wind as these form the dominant particle species in terms of the mass and momentum flux of the solar wind. More specifically, PLA's observations will allow us to construct the three-dimensional (3D) velocity distribution functions (VDFs) of the solar wind protons and determine the corresponding bulk properties by calculating the moments of the VDFs \cite{Verscharen_2019}. Vigil will record and almost instantaneously downlink in situ observations of the solar wind plasma and the interplanetary magnetic field.

Performance models are the gold-standard tool to evaluate the capability of electrostatic analyzers like PLA to determine the plasma moments. Performance models use assumed input plasma VDFs and numerically model the response of the analyzer to these input VDFs, so that the comparison between input and output parameters quantifies the expected performance of the instrument \cite<e.g.,>{Nicolaou_2014,Cara_2017,wilson_2017}.
Typically, these performance models for electrostatic analyzers use single-species Maxwellian distributions to characterise the input VDF of the incoming plasma particles \cite<e.g.,>{Verscharen_2019,Nicolaou_2020}. The Maxwellian distribution is the simplest case for the VDF since it assumes thermodynamic equilibrium with isotropic temperatures. 

In a realistic solar-wind environment, however, the plasma consists of multiple ion populations and is often non-uniform and not in a thermodynamic equilibrium state. Fast solar-wind streams, for instance, are often anisotropic \cite<e.g.,>{Marsch_1981,Marsch_2004,Bale_2009,Bourouaine_2010,Verscharen_2011}, meaning that the temperature perpendicular to the magnetic field ($T_{\perp}$) is not equal to the temperature parallel to the magnetic field ($T_{\parallel}$). Moreover, according to previous in-situ measurements, strong field-aligned proton beams are often observed with drift speeds of order or even greater than the local Alfvén speed $V_A$ \cite<e.g.,>{Marsch_1982_proton,Marsch_2006,Alterman_2018}. Moreover, there are numerous studies giving evidence that the solar wind VDFs exhibit enhanced high-energy tails which are better described with $\kappa$-distributions than with Maxwellian distributions \cite<e.g.,>{Livadiotis_2009,Livadiotis_2013,Nicolaou_2018}. In addition, $\alpha$-particles make an important contribution to the dynamics of the fast solar wind (with a mass density contribution often $\gtrsim15\%$), and $\alpha$-particles typically exhibit relative drift speeds with respect to the proton species along the magnetic field direction \cite<e.g.,>{Bame_1977,Marsch_1984,Verscharen_2015}. Some large statistical analyses of $\alpha$-particles  \cite<e.g.,>{Alterman_2018} also reflect the importance of this additional non-equilibrium species.

The goal of this study is to investigate the Vigil/PLA instrument performance under realistic solar wind conditions with non-equilibrium VDFs. We apply VDFs such as bi-Maxwellian and $\kappa$-distributions to the PLA performance model and analyse the impact of these non-equilibrium features on the determination of the plasma moments.

\section{Methodology}

PLA is an electrostatic analyzer. The basic working principle and schematic of a top-hat style electrostatic analyzer is described by \citeA{Verscharen_2019} and references therein. Below, we explain the characteristics of PLA and its performance, which are also described in detail by \citeA{Nicolaou_2020}. We build our analysis on the current baseline design for the PLA sensor. 

\begin{figure}[!h]
\centering
\noindent\includegraphics[width=\textwidth]{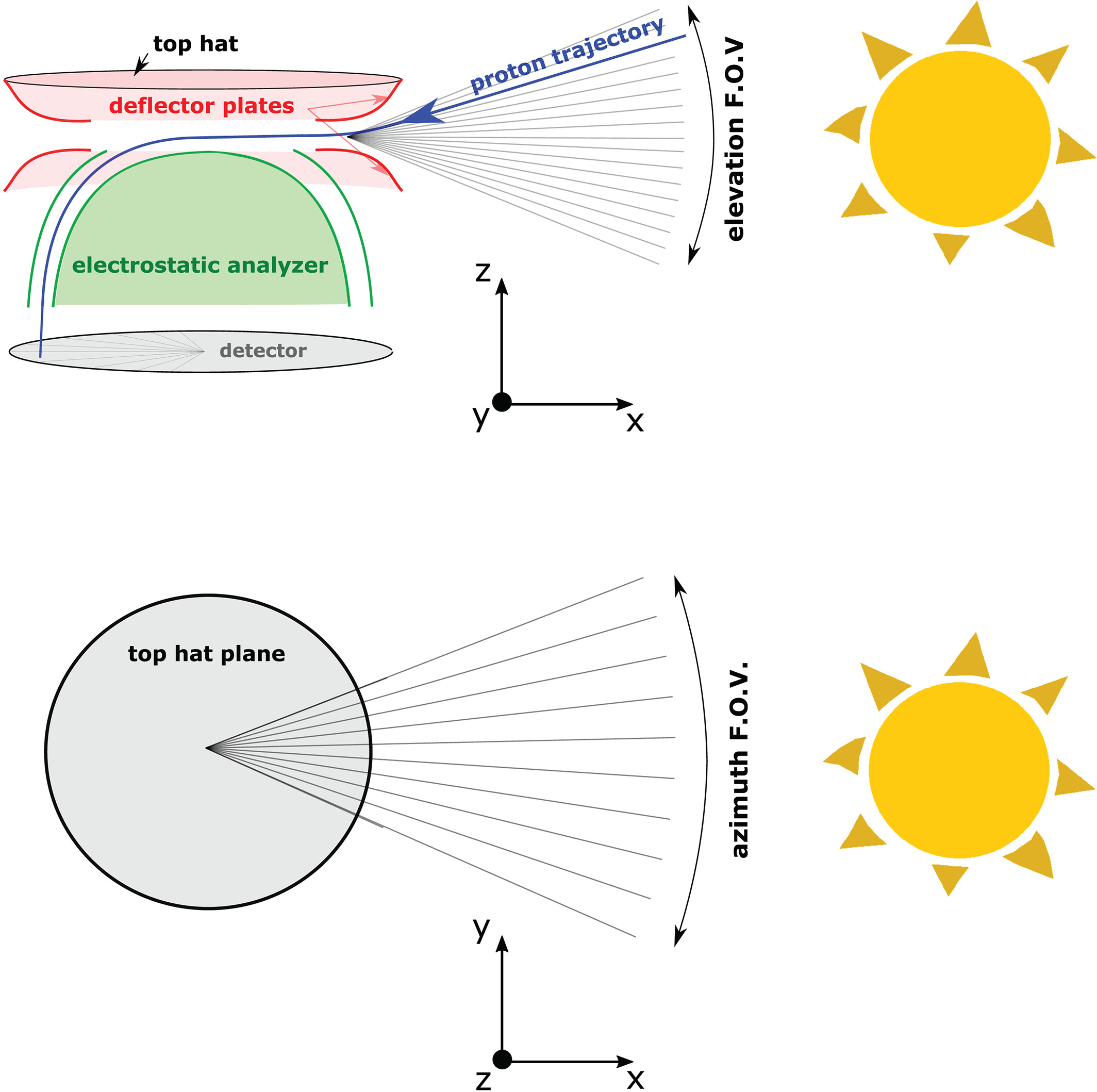}
\caption{ Schematic of the instrument design with elevation field of view (top) and azimuth field of 
view (bottom). The deflectors (red) scan the elevation angle which is the angle between the velocity 
and the sensor plane. The green 
electrostatic analyzer unit bends the trajectories of particles with specific energy-per-charge towards the detector. The azimuth angle is defined in the detector plane. From \citeA{Nicolaou_2020}.
 }
\label{deflectors}
\end{figure}

\subsection{PLA Working Principle}
 
The PLA instrument consists of an aperture deflector system, a top-hat electrostatic analyzer unit, and a detector chain. This design is built based on heritage from Solar Orbiter's SWA/EAS instrument \cite{Owen_2020}. The aperture deflectors scan through the elevation angle $\Theta$ between the velocity vector of incoming particles and the top-hat plane (same as the detector plane) as shown in Figure \ref{deflectors}. By applying an
electrostatic potential between the aperture deflectors, PLA separates particles of different
$\Theta$ from $-22.5^{\circ}$ to $+22.5^{\circ}$ in 16 discrete steps, leading to an elevation resolution of $\Delta\Theta\approx 2.8^{\circ}$. 
The elevation acceptance bandwidth depends only on the geometry of the instrument, such as the size of the aperture, and on the electrical setup of the deflection system. 

The top-hat electrostatic analyzer separates protons with specific energy-per-charge through an applied electrostatic potential between the nested semi-spherical voltage plates. By scanning through the voltage, the electrostatic analyzer measures protons between 170 and 35\,200~eV in 128 exponentially spaced energy steps, corresponding to particle speeds between about 180 and 2600~km/s. 

PLA resolves the azimuth angle $\Phi$ through 9 geometrically discrete anodes as part of its detector system, ranging from $-22.5^{\circ}$ to $+22.5^{\circ}$ with a resolution of $\Delta\Phi=5^{\circ}$. The detector uses a microchannel plate for signal amplification.  To obtain a full 3D VDF, the instrument scans through energy and elevation electrostatically and resolves the azimuth dimension geometrically and simultaneously \cite{Nicolaou_2020}. The acquisition time $\Delta\tau$ for one individual energy-elevation bin is about 0.96\,ms. Thus, the instrument completes a full VDF-scan cycle in $128 \times 16 \times \Delta\tau \approx 1.97\,\mathrm s$.

\subsection{The PLA Performance Model}

We follow a similar approach as \citeA{Nicolaou_2020} and summarize the key aspects of our performance model here \cite<see also>{wilson_2008,Nicolaou_2014,Cara_2017}. 

\subsubsection{Forward-modelling: from assumed distributions to predicted measurements}

Assuming a specific solar wind input VDF $f_{\mathrm{in}}$, the corresponding detected number of counts of particles in each $U$, $\Theta$, and $\Phi$ bin of the electrostatic analyzer at time $t$ is \cite{wuest_2007}
\begin{equation}
\label{PLA_counts_equation}
C(U, \Theta, \Phi, t)=\int\limits_{U-\frac{\Delta U}{2}}^{U+\frac{\Delta U}{2}} \int\limits_{\Theta-\frac{\Delta \Theta}{2}}^{\Theta+\frac{\Delta \Theta}{2}} \int\limits_{\Phi-\frac{\Delta \Phi}{2}}^{\Phi+\frac{\Delta \Phi}{2}} \int\limits_{t-\frac{\Delta \tau}{2}}^{t+\frac{\Delta \tau}{2}} A_{\mathrm{eff}}(u, \theta, \phi)\, f_{\mathrm{in}}(u, \theta, \phi, t^{\prime})\, u^3 \,\mathrm d u \cos \theta \,\mathrm d \theta \,\mathrm d \phi \,\mathrm d t^{\prime},
\end{equation}
where $U$ is the velocity associated with a given energy bin, and $A_{\mathrm{eff}}$ is the effective aperture of the instrument. Lower-case letters $(u,\theta,\phi)$ mark the spherical coordinates of velocity space, and $t^{\prime}$ is the time running during each acquisition bin. By applying the midpoint rule \cite<e.g.,>{Nicolaou_2020}, we approximate the number of counts of particles in each bin as
\begin{equation}
\label{PLA_counts_equation2}
C(U,\Theta,\Phi, t)\sim G(U,\Theta,\Phi)\,f_{\mathrm{in}}(U,\Theta,\Phi,t)\,U^4\,\Delta\tau,
\end{equation}
where
\begin{equation}
\label{Gfactor}
G=A_{0}\frac{\Delta U}{U}\Delta\Theta\Delta\Phi  
\end{equation}
is the geometric factor of the instrument and $A_{\mathrm{0}}$ is the effective collecting area. Although $G$ is generally dependent on energy and direction, we simplify our analysis by assuming a constant $G$-factor and set $ G= 7.8 \times 10^{-10}\,\mathrm{m^2\,sr}$ for all $(U,\Theta,\Phi)$-bins, and we assume that $f_{\mathrm{in}}$ does not vary over time during $\Delta \tau$.

An ideal instrument would count all particles arriving at the detector system. However, a real counting system cannot register all events, especially at very high particle fluxes. We introduce the dead time $t_{\mathrm{d}}$, which is the detector processing and replenishing time between two measurement events in the same anode. The detector records measurements if the separating time of two events is greater than $t_{\mathrm{d}}$. For our model, we set $t_{\mathrm{d}} = 10^{-7}\,\mathrm{s}$ based on experience with similar detector designs.

For the sampling interval $\Delta\tau$, we define the actual registered number of particles in a given anode as $C_{\mathrm{s}}$ \cite{Knoll_1989}. The total time interval over which an anode does not register events corresponds to the actually registered number of counts $C_{\mathrm{s}}$ multiplied with the individual dead time $t_{\mathrm{d}}$ of the anode. Therefore, the anode is effectively only counting particles over the time interval $\Delta \tau-C_{\mathrm{s}}t_{\mathrm{d}}$ during one acquisition step. Assuming that the plasma VDF  is constant over one acquisition step, the number of particles reaching the detection area of the anode per time remains constant according to Equation~(\ref{PLA_counts_equation2}). This allows us to relate the number of registered counts $C_{\mathrm{s}}$, accounting for the  dead time, to the detectable number of counts $C$ over a full acquisition step by:
\begin{equation}\label{counts_time}
\frac{C_{\mathrm s}}{\Delta \tau-C_{\mathrm s}t_{\mathrm d}}=\frac{C}{\Delta \tau}.
\end{equation}
Rewriting Equation~(\ref{counts_time}) leads to
\begin{equation}
\label{PLA_C_equation}
C-C_{\mathrm{s}}=C_{\mathrm{s}}C\frac{t_{\mathrm{d}}}{\Delta\tau},
\end{equation}
and thus
\begin{equation}
\label{PLA_Cs_equation}
C_{\mathrm{s}}=\frac{C}{1+\frac{t_{\mathrm{d}}}{\Delta\tau}C}.
\end{equation}

In order to account for measurement uncertainties, we assume that the counting events follow the Poisson distribution. The probability distribution of detected measurements $P(C_{\mathrm{m}})$, accounting for the error from finite counting statistics, is \cite<e.g.,>{yates_2014}
\begin{equation}\label{Poisson}
P(C_{\mathrm{m}})=e^{-C_{\mathrm{s}}}\frac{{C_{\mathrm{s}}}^{C_{\mathrm{m}}}}{C_{\mathrm{m}}!}.
\end{equation}
The resulting $C_{\mathrm m}(U,\Theta,\Phi)$ describes the predicted number of particle counts in energy, elevation, and azimuth (defined as the bin map) that PLA  detects under our assumptions for an incoming VDF $f_{\mathrm{in}}$. The variable $C_{\mathrm m}$ includes, unlike the variable $C_{\mathrm s}$, the effect of finite counting statistics on the expected number of detected counts. This is the result of our forward-modelling technique.

\subsubsection{Analysis: from predicted bin maps to moments of the VDF}

In the next step, we now analyze $C_{\mathrm m}(U,\Theta,\Phi,t)$ as if it were a detected bin map from our PLA instrument. In particular, we determine the proton moments, such as the number density, velocity, and temperature through integration. We drop the $t$-dependence of all variables at this point since we do not consider time-dependent measurement series in our analysis; instead, we treat each given $f_{\mathrm{in}}$ as an individual data point.

We first correct for the under-counted particles due to the detector's dead time. By applying the inverse of Equation~(\ref{PLA_Cs_equation}), we obtain the best estimate for the corrected counts $C_{\mathrm{out}}$ as a function of the measured number of counts $C_{\mathrm{m}}$: 
\begin{equation}
\label{PLA_Cout_equation}
C_{\mathrm{out}}=\frac{C_{\mathrm{m}}}{1-\frac{t_{\mathrm{d}}}{\Delta\tau}C_{\mathrm{m}}}.
\end{equation}
$C_{\mathrm{out}}$ is different from $C_{\mathrm{s}}$ due to the effect of finite counting statistics. Inversion of Equation~(\ref{PLA_counts_equation2}) then allows us to obtain the output VDF of the performance model as a function of $(U,\Theta,\Phi)$:
\begin{equation}
\label{PLA_fout_equation}
f_{\mathrm{out}}(U,\Theta,\Phi)=\frac{C_{\mathrm{out}}(U,\Theta,\Phi)}{GU^4\,\Delta\tau}.
\end{equation}
For an ideal instrument with infinite resolution and velocity-space coverage, $f_{\mathrm{out}}=f_{\mathrm{in}}$. However, due to finite resolution and finite counting statistics, the input and output VDFs generally do not agree exactly.

We now determine the output moments associated with $f_{\mathrm{out}}$: density $N_{\mathrm{out}}$, the components of the velocity vector $U_{i,\mathrm{out}}$, and the components of the temperature tensor $T_{\mathrm{out}}^{i,j}$ through moment integration of $f_{\mathrm{out}}$:
\begin{equation}
\label{PLA_density_equation}
N_{\mathrm{out}} = \sum_{U}^{} \sum_{\Theta}^{}\sum_{\Phi}^{}f_{\mathrm{out}}(U,\Theta,\Phi)U^2\,\Delta U\,\cos\Theta\,\Delta\Theta\,\Delta\Phi,
\end{equation}
\begin{equation}
\label{PLA_velocity_equation}
U_{i,\mathrm{out}}=\frac{1}{N_{\mathrm{out}}}\sum_{U}^{} \sum_{\Theta}^{}\sum_{\Phi}^{}U_if_{\mathrm{out}}(U,\Theta,\Phi)U^2\,\Delta U\,\cos\Theta\,\Delta\Theta\,\Delta\Phi,
\end{equation}
and 
\begin{equation}
\label{PLA_temperature_equation}
T_{\mathrm{out}}^{i,j}=\frac{1}{N_{\mathrm{out}}}\sum_{U}^{} \sum_{\Theta}^{}\sum_{\Phi}^{}m_p(w^{i,j})^2f_{\mathrm{out}}(U,\Theta,\Phi)U^2\,\Delta U\,\cos\Theta\,\Delta\Theta\,\Delta\Phi,
\end{equation}
where $w^{i,j}=\sqrt{(U_i-U_{i,\mathrm{out}})(U_j-U_{j,\mathrm{out}})}$ and $m_p$ is the proton mass. We only focus on the diagonal elements $T_{i,\mathrm{out}}=T_{\mathrm{out}}^{i,i}$ of the temperature tensor \cite<i.e, $i=j$, see>{Nicolaou_2020}.
Our approach allows us to perform an error analysis by comparing the determined moments $N_{\mathrm{out}}$, $U_{i,\mathrm{out}}$, and $T_{i,\mathrm{out}}$ of $f_{\mathrm{out}}$ with the input parameters $N_{\mathrm{in}}$, $U_{i,\mathrm{in}}$, and $T_{i,\mathrm{in}}$ of $f_{\mathrm{in}}$.

Most previous models assume an istropic Maxwellian distribution for $f_{\mathrm{in}}$ to describe the solar wind in thermal equilibrium \cite<e.g.,>{Verscharen_2019,Nicolaou_2020}. In this case, the proton VDF is given by
\begin{equation}
\label{max_equation_PLA}
f_p(\vec{u})=N_{\mathrm{in}}\left(\frac{m_p}{2 \pi k_B T_{\mathrm{in}}}\right)^{3/2} \exp \left(-\frac{m_p(\vec{u}-\Vec{U}_{\mathrm{in}})^2}{2 k_B T_{\mathrm{in}}}\right),
\end{equation}
where $T_{\mathrm{in}}$ is the scalar proton temperature and $k_B$ is the Boltzmann constant. This case is examined in detail by \citeA{Nicolaou_2020}. In this study, we consider non-equilibrium distributions \cite<e.g.,>{Hellinger_2006,kasper_2006,Marsch_2006,Verscharen_2019} for $f_{\mathrm{in}}$ and study their impact on the determination of the plasma moments.

\subsection{Temperature anisotropy}
\label{section:method:Apply temperature anisotropy}

We model an anisotropic plasma through a bi-Maxwellian input distribution in cylindrical velocity space:
\begin{equation}
\label{bimax_equation}
f_b(v_{\perp},v_{\parallel})=\frac{N_{\mathrm{in}}}{\pi^{3/2} w_{\perp j}^2 w_{\| j}} \exp \left(-\frac{v_{\perp}^2}{w_{\perp j}^2}-\frac{\left(v_{\|}-U_{\| in}\right)^2}{w_{\| j}^2}\right),
\end{equation}
where $w_{\perp}=\sqrt{2k_BT_{\perp}/m_p}$ and $w_{\parallel}=\sqrt{2k_BT_{\parallel}/m_p}$ are the thermal velocities in the directions perpendicular and parallel to the magnetic field. 
We transform $f_b$ into a three-dimensional Cartesian coordinate system $(x,y,z)$ and assume that the spacecraft--Sun axis is aligned with the $x$-direction. The azimuth plane of the detector then corresponds to the $xy$-plane. For simplicity, we assume that the magnetic field is parallel to the $x$-direction so that parallel velocities are aligned with the $x$-direction as well, and perpendicular velocity components lie in the $yz$-plane of the instrument. In principle, the expected magnetic field direction is statistically quasi-random around the mean associated with the Parker spiral direction. The variability of the magnetic-field direction primarily impacts the temperature moments and their uncertainties. However, a detailed analysis of the dependence of our results on the direction of the magnetic field is beyond the scope of this work. In this case, Equation~(\ref{max_equation_PLA}) yields the Cartesian VDF
\begin{equation}
\label{bi_max_equation2}
f_{b}^{\prime}(\vec{u})=\frac{N_{\mathrm{in}}}{\pi^{3/2} w_x w_y w_z} \exp \left(-\frac{\left(u_x-U_{x,\mathrm{in}}\right)^2}{w_x^2}-\frac{\left(u_y-U_{y,\mathrm{in}}\right)^2}{w_y^2}-\frac{\left(u_z-U_{z,\mathrm{in}}\right)^2}{w_z^2}\right),
\end{equation}
where $w_x$, $w_y$, and $w_z$ are the anisotropic thermal velocity components so that $w_i=\sqrt{2k_BT_{i,\mathrm{in}}/m_p }$. In the fast solar wind, the perpendicular temperature of the protons is often greater than their parallel temperature, yet the distribution is still gyrotropic. To reflect this geometry, we set $w_y=w_z=2w_x$ in our model.

\subsection{Proton and $\alpha$-particle beams}
\label{section:method:Proton and alpha-particle beams}

For the sake of simplicity, we assume a Maxwellian proton beam. We add the proton beam distribution function ($f_{\mathrm{beam}}$) to the distribution function of the proton core ($f_{\mathrm{core}}$), yet with a different density $N_{\mathrm{in}}^{\prime}$,  bulk velocity $U_{x,\mathrm{in}}^{\prime}$, and temperature $T_{i,\mathrm{in}}^{\prime}$ according to Equation~(\ref{max_equation_PLA}). We then apply the sum of both distributions to our performance model as a new input distribution function: $f_{\mathrm{in}} = f_{\mathrm{beam}} + f_{\mathrm{core}}$.  

The $\alpha$-particles cannot be simply added to the proton distribution as they represent a separate species of particles. However, an electrostatic analyzer like PLA cannot distinguish between different species as all particles with the same energy-per-charge appear at the same energy bin so that an unambiguous separation by charge, mass, and velocity is not possible with this detector design. Since an $\alpha$-particle has four times the mass of a proton and carries twice the charge of a proton, the energy per charge ($E_{\alpha}/q_{\alpha}$) of an $\alpha$-particle with the same speed as a proton is twice the energy-per-charge ($E_p/q_p$) of that proton. Consequently, the speed of an $\alpha$-particle that is detected at the same energy-per-charge as a proton (i.e., when $E_{\alpha}/q_{\alpha}=E_p/q_p$) can be assumed by a factor $\sqrt{2}$ smaller than the speed of the proton \cite{Nicolaou_2022}.

In order to examine the impact of  $\alpha$-particles on our measurement of the solar wind protons, we add $\alpha$-particles ($C_{\alpha}$) to the expected number of proton counts ($C_{p}$) according to Equation~(\ref{PLA_counts_equation2}) at their corresponding $E_{\alpha}/q_{\alpha}$ to interpret them (wrongfully, but realistically for an electrostatic analyzer) as protons. Hence the new expected counts fulfill
\begin{equation}\label{Calpha}
C(U,\Phi,\Theta)=G f_{\mathrm{in},p}(U,\Theta,\Phi)\,U^4\,\Delta \tau+\frac{G}{4} f_{\mathrm{in},\alpha}(U/\sqrt 2,\Theta,\Phi)\,U^4\,\Delta \tau,
\end{equation}
where $f_{\mathrm{in},p}$ is the input VDF of the protons and $f_{\mathrm{in},\alpha}$ is the input VDF of the $\alpha$-particles with the moments $N_{\mathrm{in}}^{\prime\prime}$, $U_{x,\mathrm{in}}^{\prime\prime}$, and $T_{i,\mathrm{in}}^{\prime\prime}$. 
We then quantify the effects produced by these $\alpha$-particles on the proton moment determination by analysing $C$ from Equation~(\ref{Calpha}) as if this function were built up by protons only. 
We introduce the definition of our input parameters in Table \ref{input_bulk_parameters_table}.

\begin{table}
\caption{Summary of the input bulk parameters for the plasma.}
\label{input_bulk_parameters_table}
 \centering
 \begin{tabular}{l c c c}
 \hline
   & Number density & Bulk velocity & Temperature \\
 \hline
   Core  & $N_{\mathrm{in}}$ &  $U_{x,\mathrm{in}}$ & $T_{i,\mathrm{in}}$   \\
    Beam  & $N_{\mathrm{in}}^{\prime}$ &$U_{x,\mathrm{in}}^{\prime}$ & $T_{i,\mathrm{in}}^{\prime}$  \\
    $\alpha$-particles & $N_{\mathrm{in}}^{\prime\prime}$ & $U_{x,\mathrm{in}}^{\prime\prime}$ & $T_{i,\mathrm{in}}^{\prime\prime}$\\
 \hline
 \end{tabular}
 \end{table}

\section{Results}

In this section, we present the results of our instrument performance model. The raw output of the model consists of maps of particle counts in energy, elevation, and azimuth bins.  
To demonstrate the accuracy of the instrument's measurements, we calculate the ratios of the calculated moments over the input moments. We define the accuracy of the density as $\langle N_{\mathrm{out}}\rangle/N_{\mathrm{in}}$, the accuracy of the components of the bulk velocity as $\langle U_{i,\mathrm{out}}\rangle/U_{i,\mathrm{in}}$, and the accuracy of the components of the temperature as $\langle T_{i,\mathrm{out}}\rangle/T_{i,\mathrm{in}}$, where $i\in (x,y,z)$ and $\langle\cdot\rangle$ describes the mean of ten evaluations of the performance model. These evaluations differ due to the statistical process introduced in Equation~(\ref{Poisson}).

\subsection{Proton temperature anisotropy}
\label{section:result:Bi-Maxwellian distribution function}

\begin{figure}[!h]
\centering
\noindent\includegraphics[width=\textwidth]{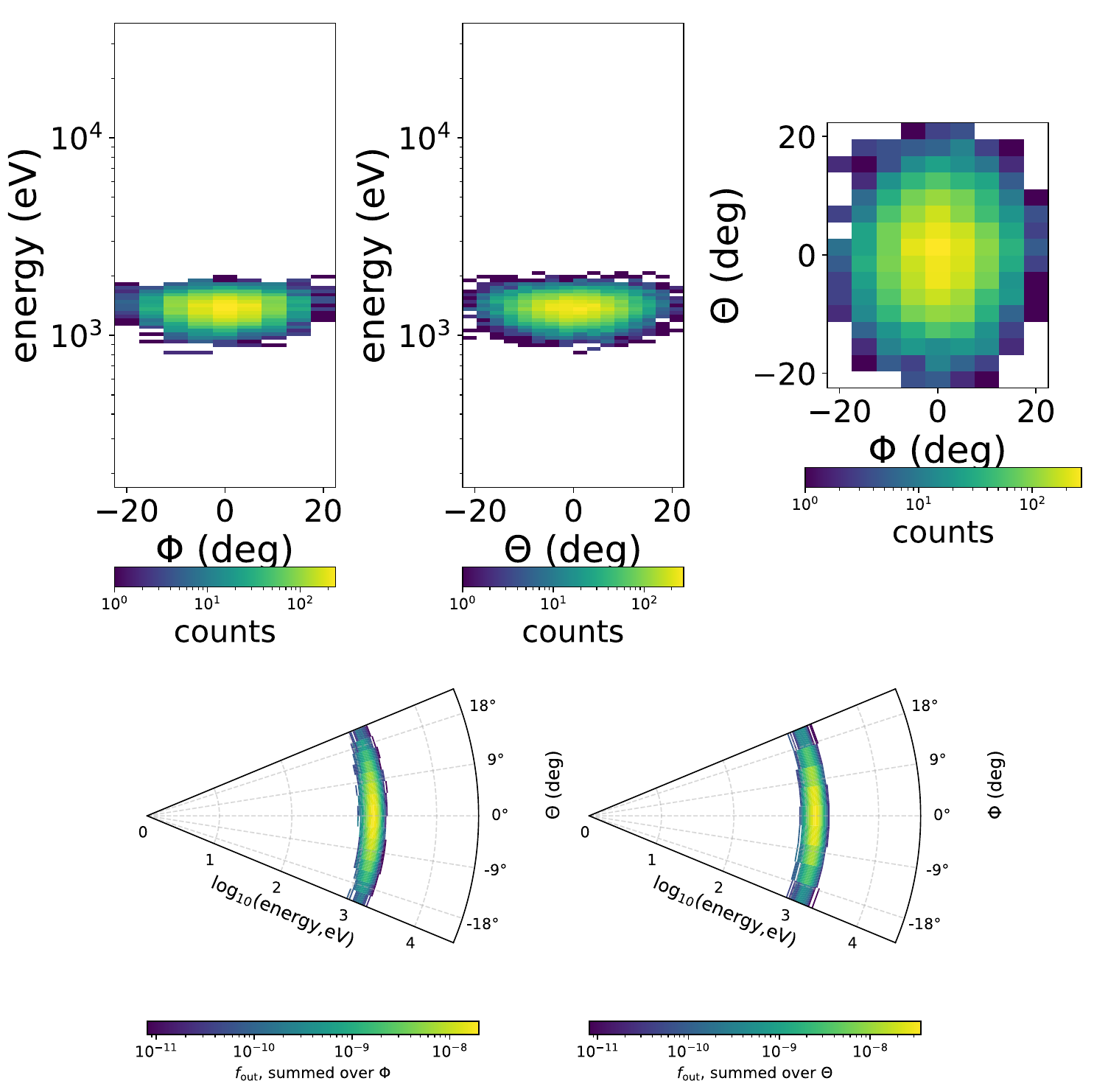}
\caption{Result of our performance model for an anisotropic input plasma with a bi-Maxwellian proton distribution. We use $N_{\mathrm{in}}=10\,\mathrm{cm}^{-3}$, $U_{x,\mathrm{in}} = 500\,\mathrm{km/s}$, $U_{y,\mathrm{in}} = U_{z,\mathrm{in}} = 0\,\mathrm{km/s}$, and $T_{y,\mathrm{in}}=T_{z,\mathrm{in}}=4T_{x,\mathrm{in}}=40\,\mathrm{eV}$. Top: Count maps in energy, azimuth, and elevation. Bottom: Output VDF $f_{\mathrm{out}}$ in energy-angle space, summed over the other angle. }
\label{bi-max_simulation}
\end{figure}

In Figure \ref{bi-max_simulation}, we show our simulation results for the anisotropic solar wind case using a bi-Maxwellian input according to Equation~(\ref{bi_max_equation2}) with $T_{y,\mathrm{in}}=T_{z,\mathrm{in}}=4T_{x,\mathrm{in}}$. The top panels show maps of  $C_m$ as a function of energy and angle, while the bottom panels show the reconstructed solar wind VDF $f_{\mathrm{out}}$ as a function of energy, azimuth, and elevation. The distribution has its peak at $\Theta\sim0^{\circ}$ and $\Phi\sim 0^{\circ}$, which is the direction along the bulk velocity vector we assume in our model. The vertical axis in the first two panels shows the energy of the detected particles.  We recognize the temperature anisotropy as an elongation in the azimuth and elevation directions (i.e, the directions perpendicular to the assumed magnetic field) compared to the Maxwellian case, which would appear as a symmetric distribution \cite<see>{Nicolaou_2020}.

Table~\ref{bi_max_table} shows a comparison between the input and output plasma moments for this case. We compare all three components of the temperature ($x$, $y$, and $z$). In this case of an anisotropic plasma, the instrument obtains
measurements that lead to accurate calculations of all moments within 5\% of the input moments.

 \begin{table}[!h]
 \caption{ Input and output moments for the anisotropic bi-Maxwellian proton case.}
 \label{bi_max_table}
 \centering
 \begin{tabular}{l c c c c c c c c }
 \hline
  Moment & $N(\mathrm{cm}^{-3})$ & $U_x(\mathrm{km/s})$ & $U_y(\mathrm{km/s})$ & $U_z(\mathrm{km/s})$ & $T_x(\mathrm{eV})$ & $T_y(\mathrm{eV})$ & $T_z(\mathrm{eV})$ & $T_{avg}(eV)$\\
 \hline
   Input  &10 & 500 &0 &0 &10 &40 &40 &30 \\
    Output  &10.2 & 500.0 &0.1 &0.0 &9.9 &39.5 &39.4 & 29.6 \\
 \hline
 \end{tabular}
 \end{table}

\begin{figure}[!h]
\centering
\noindent\includegraphics[width=\textwidth]{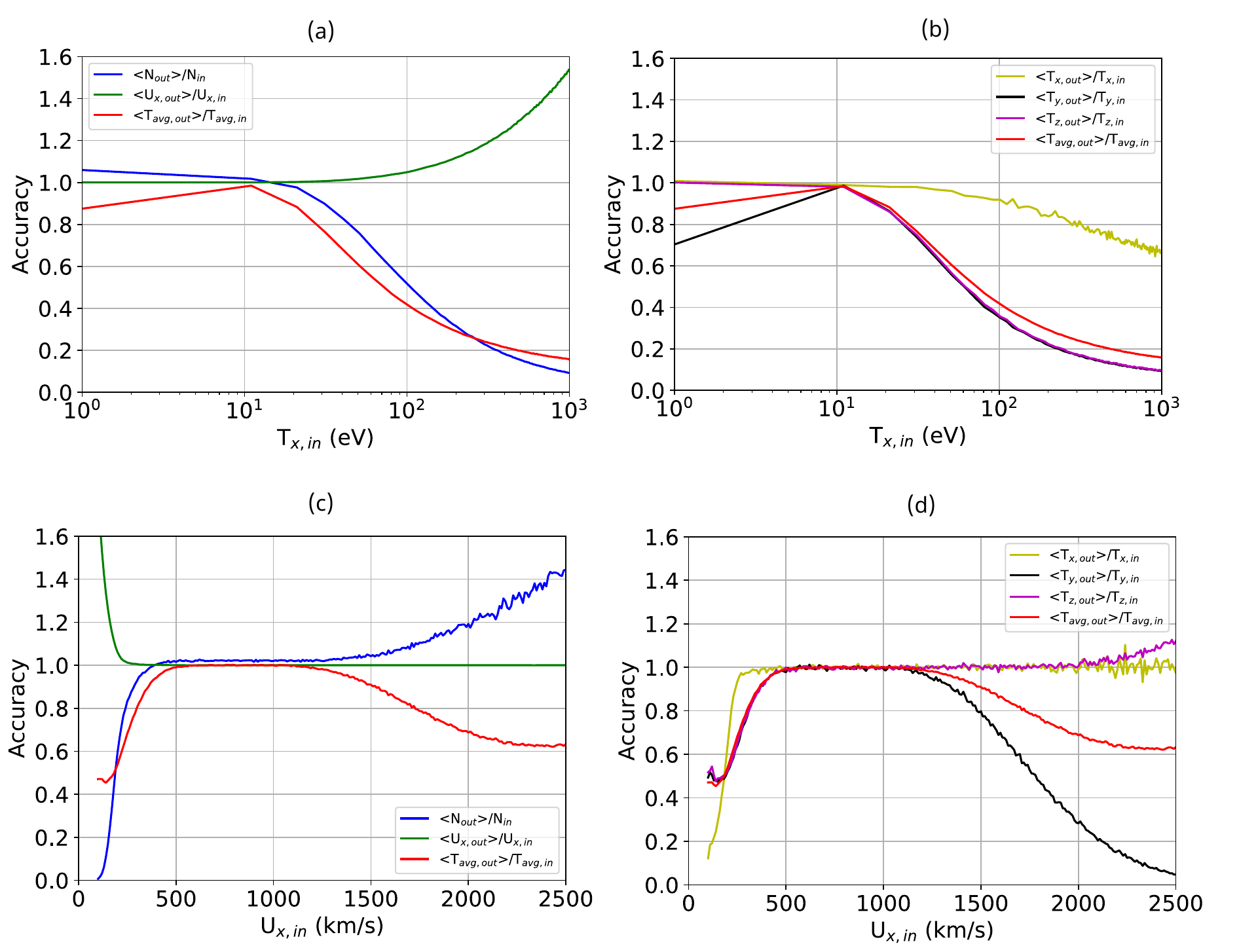}
\caption{Measurement accuracy for an anisotropic input plasma with a bi-Maxwellian proton VDF with $N_{\mathrm{in}}=10\,\mathrm{cm}^{-3}$, $U_{y,\mathrm{in}}=U_{z,\mathrm{in}}=0\,\mathrm{km/s}$, and $T_{y,\mathrm{in}}=T_{z,\mathrm{in}}=4T_{x,\mathrm{in}}$.
 (a) Accuracy of the moments $N_{\mathrm{out}}$, $U_{x,\mathrm{out}}$, and $T_{\mathrm{avg,out}}$ as a function of $T_{x,\mathrm{in}}$ for  $U_{x,\mathrm{in}}=500\,\mathrm{km/s}$. (b) Accuracy of the temperatures $T_{x,\mathrm{out}}$, $T_{y,\mathrm{out}}$, $T_{z,\mathrm{out}}$, and $T_{\mathrm{avg,out}}$ as a function of $T_{x,\mathrm{in}}$ for  $U_{x,\mathrm{in}}=500\,\mathrm{km/s}$. (c) Accuracy of $N_{\mathrm{out}}$, $U_{x,\mathrm{out}}$, and $T_{\mathrm{avg,out}}$ as a function of $U_{x,\mathrm{in}}$ for  $T_{x,\mathrm{in}}=10\,\mathrm{eV}$.  (d) Accuracy of the temperatures $T_{x,\mathrm{out}}$, $T_{y,\mathrm{out}}$,  $T_{z,\mathrm{out}}$, and  $T_{\mathrm{avg,out}}$ as a function of $U_{x,\mathrm{in}}$ for  $T_{x,\mathrm{in}}=10\,\mathrm{eV}$.}
\label{bi-max_analysis}
\end{figure}

In Figure~\ref{bi-max_analysis}a, we show the dependence of the accuracy of the moments on the input parallel temperature $T_{x,\mathrm{in}}$. We use $T_{y,\mathrm{in}} = T_{z,\mathrm{in}} = 4T_{x,\mathrm{in}}$. The input average temperature $T_{\mathrm{avg},\mathrm{in}}$ is defined as $(T_{x,\mathrm{in}}+T_{y,\mathrm{in}}+T_{z,\mathrm{in}})/3$. We calculate the output average temperature $T_{\mathrm{avg},\mathrm{out}}$ likewise. As $T_{x,\mathrm{in}}$ increases, the accuracy of each measurement approaches unity (i.e, a good agreement between input and output) and is flat for $T_{x,\mathrm{in}}\sim 10\,\mathrm{eV}$. As $T_{x,\mathrm{in}}$ increases further, the output velocity becomes overestimated, and the output density and average temperature values are underestimated.

Figure \ref{bi-max_analysis}b shows the accuracy of the three temperature components and their average. In cold solar wind (i.e., $T_{x,\mathrm{in}}\lesssim 10\,\mathrm{eV}$), the accuracy of the temperature measurement varies in the different directions. The most accurate results are obtained when the input parallel temperature is around 10\,eV. As the temperature increases, the accuracy of the temperature measurement decreases in all components, but more slowly in the $x$-component. In addition, the trends in the perpendicular $y$- and $z$-components are almost identical.

Figure~\ref{bi-max_analysis}c shows the accuracy plot as a function of $U_{x,\mathrm{in}}$. All moments are measured incorrectly for very slow solar wind ($U_{x,\mathrm{in}} < 300\,\mathrm{km/s} $). With increasing speed, the 
measurements are more accurate, especially between $500\,\mathrm{km/s}$ and $1250\,\mathrm{km/s}$. For extremely fast solar wind, the measured number density becomes overestimated and the measured temperature underestimated.  

Figure~\ref{bi-max_analysis}d shows the accuracy of the individual temperature components as a function of  $U_{x,\mathrm{in}}$. The measurements $T_{x,\mathrm{out}}$ and $T_{z,\mathrm{out}}$  are more accurate than the measurement $T_{y,\mathrm{out}}$ at large $U_{x,\mathrm{in}}$. We also note a difference in the $x$- and $z$-components of the temperature at very high speeds (10\% at $2500\,\mathrm{km/s}$). Moreover, for very slow solar wind ($<300\,\mathrm{km/s}$), the measurement accuracy of all temperature components breaks down.
Similar to the isotropic Maxwellian case \cite{Nicolaou_2020}, the moment measurement does not depend on the input density over a range of $N_{\mathrm{in}}$ from about 2 to 1100\,cm$^{-3}$.

\subsection{Proton beams and $\alpha$-particles}

We focus our analysis on two main cases: (i) a solar wind distribution consisting of a proton core and a proton beam, and (ii) a solar wind distribution consisting of a proton core, a proton beam, and drifting $\alpha$-particles. We study the separate impact of $\alpha$-particles without the presence of a proton beam in \ref{appendix_alphas}. 

\subsubsection{Effect of proton beams}
\label{section:result:Distribution with proton beam}

\begin{figure}[!h]
\centering
\noindent\includegraphics[width=\textwidth]{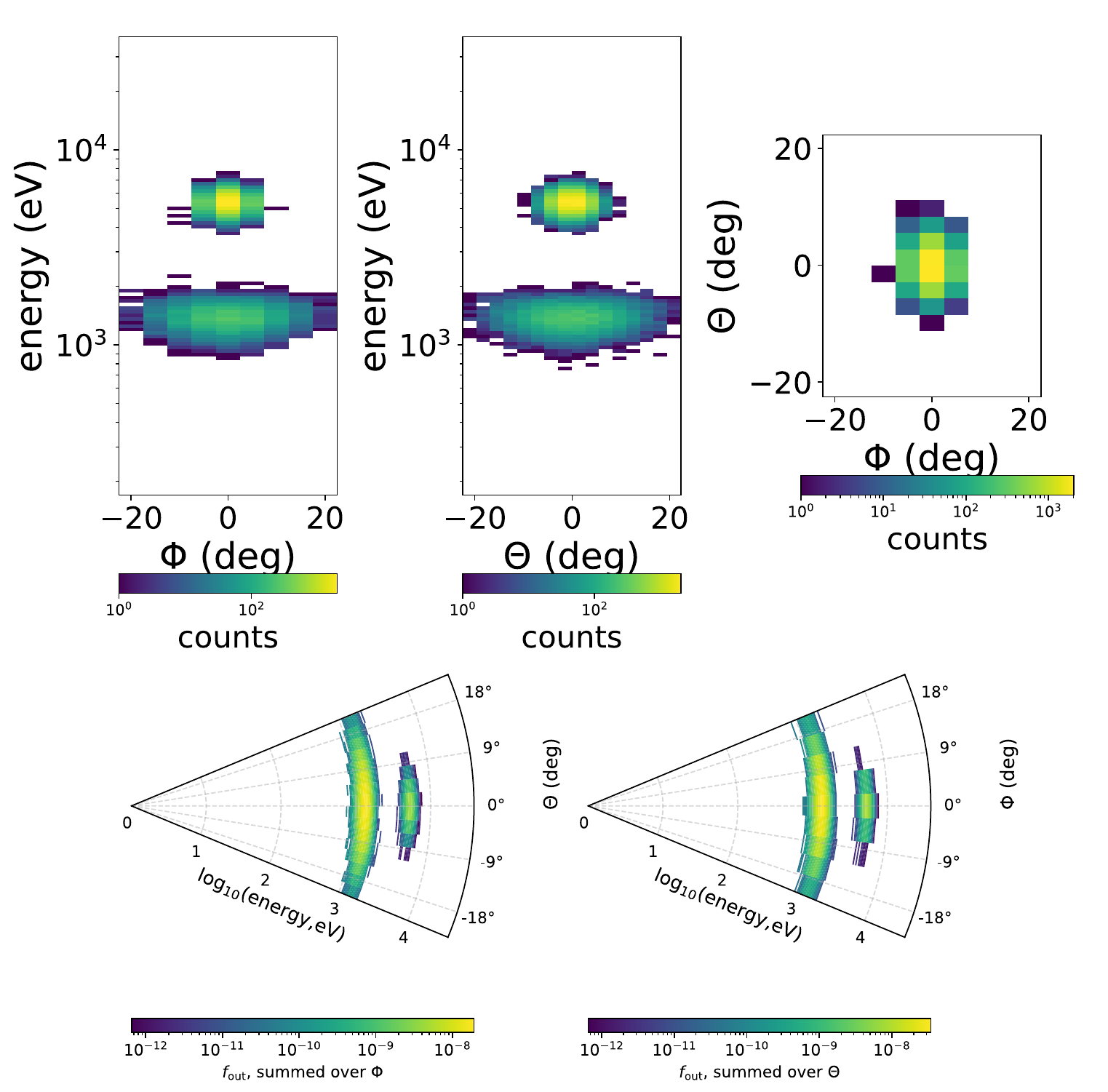}
\caption{
Result of our performance model for input plasma with a bi-Maxwellian proton core and an isotropic Maxwellian proton beam. The core input parameters are: $N_{\mathrm{in}}=10\,\mathrm{cm}^{-3}$, $U_{x,\mathrm{in}} = 500\,\mathrm{km/s}$, and $T_{y,\mathrm{in}}=T_{z,\mathrm{in}}=4T_{x,\mathrm{in}}=40\,\mathrm{eV}$. The beam input parameters are: ${N_{\mathrm{in}}}^{\prime}=5\,\mathrm{cm}^{-3}$, ${U_{x,\mathrm{in}}}^{\prime} = 1000\,\mathrm{km/s}$, and ${T_{y,\mathrm{in}}}^{\prime}={T_{z,\mathrm{in}}}^{\prime}={T_{x,\mathrm{in}}}^{\prime}=20\,\mathrm{eV}$. Top: Count maps in energy, azimuth, and elevation. Bottom: Output VDF $f_{\mathrm{out}}$ in energy-angle space, summed over the other angle. 
}
\label{proton_beam_simulation}
\end{figure}

In this section, we study the effect of adding a proton beam to a proton core on our moment analysis. We assume that the proton core is anisotropic (as in Section \ref{section:method:Apply temperature anisotropy}), and we then add a hotter ($T_{\mathrm{in}}^{\prime}=20\,\mathrm{eV}$) and faster ($U_{x,\mathrm{in}}^{\prime}=1000\,\mathrm{km/s}$) proton beam with isotropic temperature. As shown in Figure~\ref{proton_beam_simulation}, the proton beam is visible at higher energies than the proton core. 

Table~\ref{proton_beam_table} provides the input and output moments of the proton core and proton beam. The expected output number density is the sum of the individual input number densities of the proton core and proton beam: 
\begin{equation}
N_{\mathrm{tot}}=N_{\mathrm{in}}+N_{\mathrm{in}}^{\prime}=15\,\mathrm{cm}^{-3}.
\end{equation}
The expected output bulk velocity of the total proton distribution (core and beam combined) is the density-weighted average of the proton core and proton beam input bulk velocities (i.e., the proton center-of-mass velocity): 
\begin{equation}
U_{\mathrm{tot}}=\frac{N_{\mathrm{in}}U_{x,\mathrm{in}}+N_{\mathrm{in}}^{\prime}U_{x,\mathrm{in}}^{\prime}}{N_{\mathrm{tot}}}\approx 667\,\mathrm{km/s}.
\end{equation}

The temperature of the total proton distribution in the proton bulk-speed frame is given as the second moment of the full distribution:
\begin{multline}\label{total_temp}
T_{\mathrm{tot}}=\frac{1}{{N_{\mathrm{in}}+N_{\mathrm{in}}^{\prime}}}\left[N_{\mathrm{in}}T_{avg,\mathrm{in}}+N_{\mathrm{in}}^{\prime}T_{\mathrm{in}}^{\prime}\right.\\
\left.+N_{\mathrm{in}}\frac{m_{ p}}{3k_{ B}}\left(U_{\mathrm{in}}-U_{\mathrm{tot}}\right)^2+N_{\mathrm{in}}^{\prime}\frac{m_{ p}}{3k_{ B}}\left(U_{\mathrm{in}}^{\prime}-U_{\mathrm{tot}}\right)^2\right]
\end{multline}
where $T_{avg,\mathrm{in}} = (T_{x,\mathrm{in}}+T_{y,\mathrm{in}}+T_{z,\mathrm{in}})/3$. Using our input parameters in Equation~(\ref{total_temp}), gives $T_{\mathrm{tot}}=220\,\mathrm{eV}$, which corresponds to the measured output temperature $T_{\mathrm{out}}$.

According to Table~\ref{proton_beam_table}, all proton moments are close to their expected input values. 

 \begin{table}[!h]
 \caption{
 Input and output moments for the anisotropic bi-Maxwellian proton core and the isotropic Maxwellian proton beam.}
 \label{proton_beam_table}
 \centering
 \begin{tabular}{l l c c c c c }
 \hline
    & Moment & $N(\mathrm{cm^{-3}})$ & $U_x(\mathrm{km/s})$ & $U_y(\mathrm{km/s})$ & $U_z(\mathrm{km/s})$ & $T_x(\mathrm{eV})$ \\
 \hline
   Core  & Input  &10 & 500 &0 &0 &10 \\
   Beam  & Input  &5 & 1000 &0 &0 &20 \\
    Total  & Output  &15.4 & 667.9 &-0.2 &0.2 &220.2 \\
 \hline
 \end{tabular}
 \end{table}

\begin{figure}[!h]
\centering
\noindent\includegraphics[width=\textwidth]{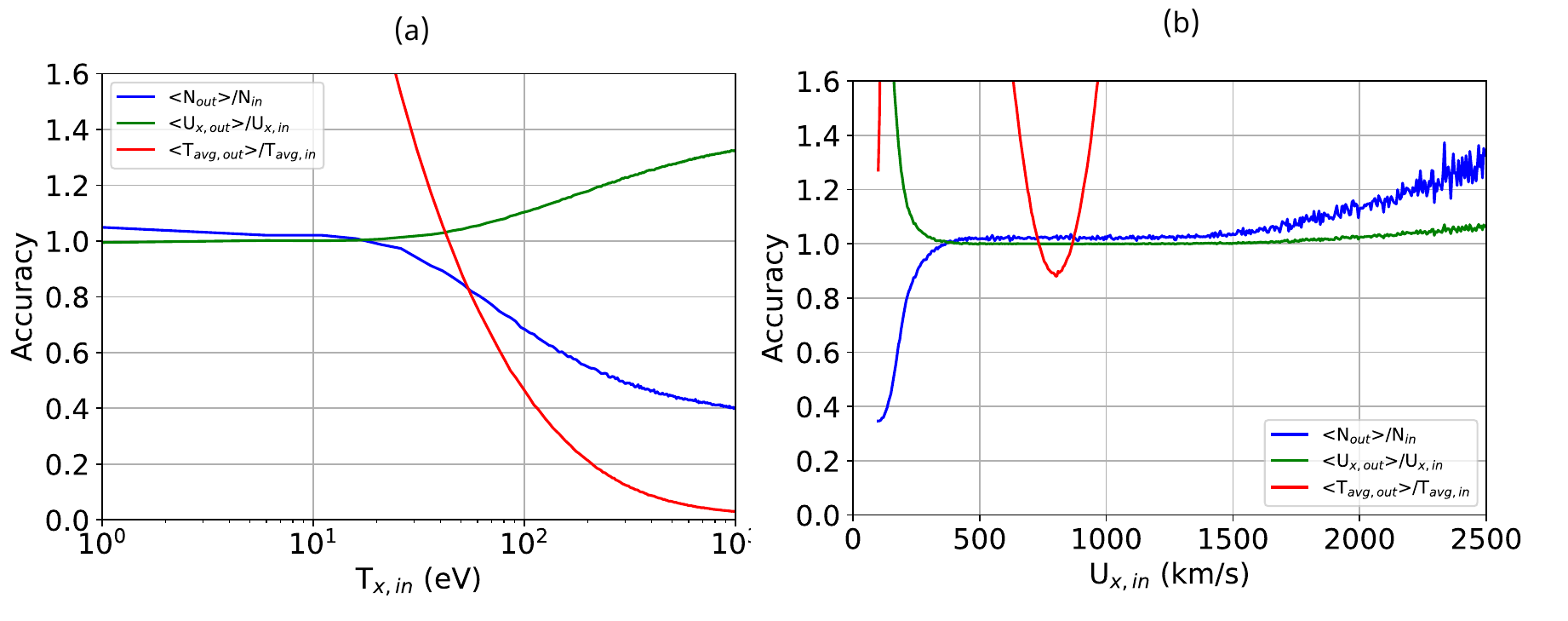}
\caption{
Measurement accuracy for an input plasma with a bi-Maxwellian proton core distribution with $N_{\mathrm{in}}=10\,\mathrm{cm}^{-3}$, $U_{y,\mathrm{in}}=U_{z,\mathrm{in}}=0\,\mathrm{km/s}$, and $T_{y,\mathrm{in}}=T_{z,\mathrm{in}}=4T_{x,\mathrm{in}}$, and an isotropic-Maxwellian proton beam with $N_{\mathrm{in}}^{\prime}=5\,\mathrm{cm}^{-3}$, $U_{x,\mathrm{in}}^{\prime} = 800\,\mathrm{km/s}$, and ${T_{y,\mathrm{in}}}^{\prime}={T_{z,\mathrm{in}}}^{\prime}={T_{x,\mathrm{in}}}^{\prime}=20\,\mathrm{eV}$. Left: Accuracy of $N_{\mathrm{out}}$ and $U_{x,\mathrm{out}}$ as a function of $T_{x,\mathrm{in}}$ for  $U_{x,\mathrm{in}}=500\,\mathrm{km/s}$. Right: Accuracy of $N_{\mathrm{out}}$ and $U_{x,\mathrm{out}}$ as a function of $U_{x,\mathrm{in}}$ for $T_{x,\mathrm{in}}=10\,\mathrm{ev}$.}
\label{proton_beam_analysis}
\end{figure}

Figure \ref{proton_beam_analysis}a shows the accuracy plot for the number density, bulk velocity, and temperature measurements under different input temperatures of the proton core. The number density curve gradually approaches unity at low temperatures and then begins to drop as the temperature increases, further finally reaching a minimum of 0.4 when  $T_{x,\mathrm{in}}\approx 1000\,\mathrm{eV}$. Although this overall trend is very similar to the trend before adding the proton beam, the accuracy of the density and the bulk velocity in the high-temperature region has improved (about 30 percentage points better at $T_{x,\mathrm{in}}=1000\,\mathrm{eV}$ ) compared to the results shown in Figure~\ref{bi-max_analysis}a.  

Figure \ref{proton_beam_analysis}b shows the accuracy plot for the number density, velocity, and temperature with different $U_{x,\mathrm{in}}$ after adding the proton beam to the model. The overall trends of the number density and velocity curves are similar to our previous results for the bi-Maxwellian case without proton beam. However, for slow solar wind, the velocity is overestimated by a factor of more than 2.25, and there are also overestimations in very fast solar wind. The measurement of the total number density is more accurate than without adding a proton beam for extreme solar wind speeds. The temperature curve shows that a reliable measurement of the core temperature is not possible due to the presence of the beam in the shown parameter combination.

The moments $N_{\mathrm{out}}$ and $U_{x,\mathrm{out}}$ for the total distribution depend only slightly on the selected $N_{\mathrm{in}}$ (not shown).

\subsubsection{Effect of proton beams and $\alpha$-particles}
\label{section:result:Distribution with alpha particles}

We now include $\alpha$-particles, assuming that their velocities follow an isotropic Maxwellian distribution function, their number density is $N_{\mathrm{in}}^{\prime\prime}=0.04/N_{\mathrm{in}}$  \cite{Alterman_2019}, and their bulk velocity $U_{i,\mathrm{in}}^{\prime\prime}$ is exactly the same as that of the proton core. In addition, we retain a faster proton beam (with $U_{x,\mathrm{in}}^{\prime}= 1000$\,km/s) to distinguish visually between $\alpha$-particles and the proton beam in our distribution plots. 

\begin{figure}[!h]
\centering
\noindent\includegraphics[width=\textwidth]{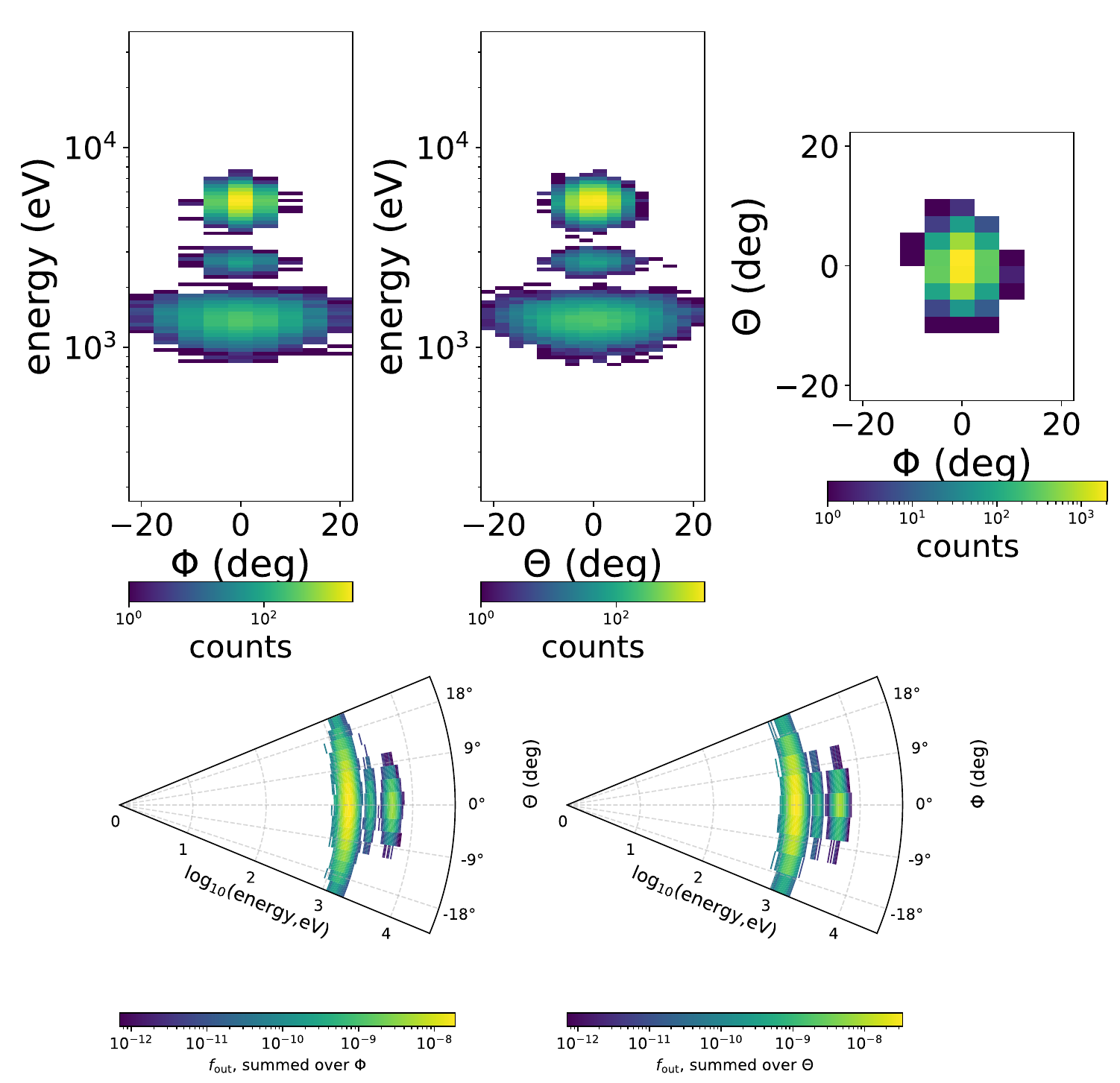}
\caption{
Result of our performance model for a plasma with a bi-Maxwellian proton core, an isotropic proton beam, and an isotropic population of $\alpha$-particles (4$\mathrm{\%}$ of the proton core density) with input proton core moments: $N_{\mathrm{in}}=10\,\mathrm{cm}^{-3}$, $U_{x,\mathrm{in}} = 500\,\mathrm{km/s}$, and $T_{y,\mathrm{in}}=T_{z,\mathrm{in}}=4T_{x,\mathrm{in}}=40\,\mathrm{eV}$. The input proton beam moments are: ${N_{\mathrm{in}}}^{\prime}=5\,\mathrm{cm}^{-3}$, ${U_{x,\mathrm{in}}}^{\prime} = 1000\,\mathrm{km/s}$, and ${T_{y,\mathrm{in}}}^{\prime}={T_{z,\mathrm{in}}}^{\prime}={T_{x,\mathrm{in}}}^{\prime}=20\,\mathrm{eV}$. The input $\alpha$-particle moments are ${N_{\mathrm{in}}}^{\prime\prime}=0.4\,\mathrm{cm}^{-3}$, ${U_{x,\mathrm{in}}}^{\prime\prime } = 500\,\mathrm{km/s}$, and ${T_{y,\mathrm{in}}}^{\prime\prime}={T_{z,\mathrm{in}}}^{\prime\prime}=4{T_{x,\mathrm{in}}}^{\prime\prime }=40\,\mathrm{eV}$. Top: Count maps in energy, azimuth, and elevation. Bottom: Output VDF $f_{\mathrm{out}}$ in energy-angle space, summed over the other angle. 
}
\label{alpha_simulation}
\end{figure}

Figure~\ref{alpha_simulation} shows the result of our performance model after adding the proton beam and $\alpha$-particles into the model. The $\alpha$-particles appear as an additional species with higher energies than the proton core, for the reasons discussed in Section~\ref{section:method:Proton and alpha-particle beams}. According to the data in Table~\ref{proton_beam_alpha_table}, $N_{\mathrm{out}}$ and $U_{x,\mathrm{out}}$ are affected by the $\alpha$-particles compared to the case without $\alpha$-particles.

\begin{table}[!h]
 \caption{ 
 Input and output moments for the anisotropic bi-Maxwellian proton core distribution with an isotropic Maxwellian proton beam and $\alpha$-particles.
 }
 \label{proton_beam_alpha_table}
 \centering
 \begin{tabular}{l l c c c c c }
 \hline
    & Moment & $N(\mathrm{cm^{-3}})$ & $U_x(\mathrm{km/s})$ & $U_y(\mathrm{km/s})$ & $U_z(\mathrm{km/s})$ & $T_x(\mathrm{eV})$ \\
 \hline
   Core  & Input  &10 & 500 &0 &0 &10 \\
   Beam  & Input  &5 & 1000 &0 &0 &20 \\
   $\alpha$-particles  & Input  &0.4 & 500 &0 &0 &10 \\
   Total  & Output  &15.6 & 668.5 &0.1 &0.1 &217.1 \\
 \hline
 \end{tabular}
 \end{table}

\begin{figure}[!h]
\centering
\noindent\includegraphics[width=\textwidth]{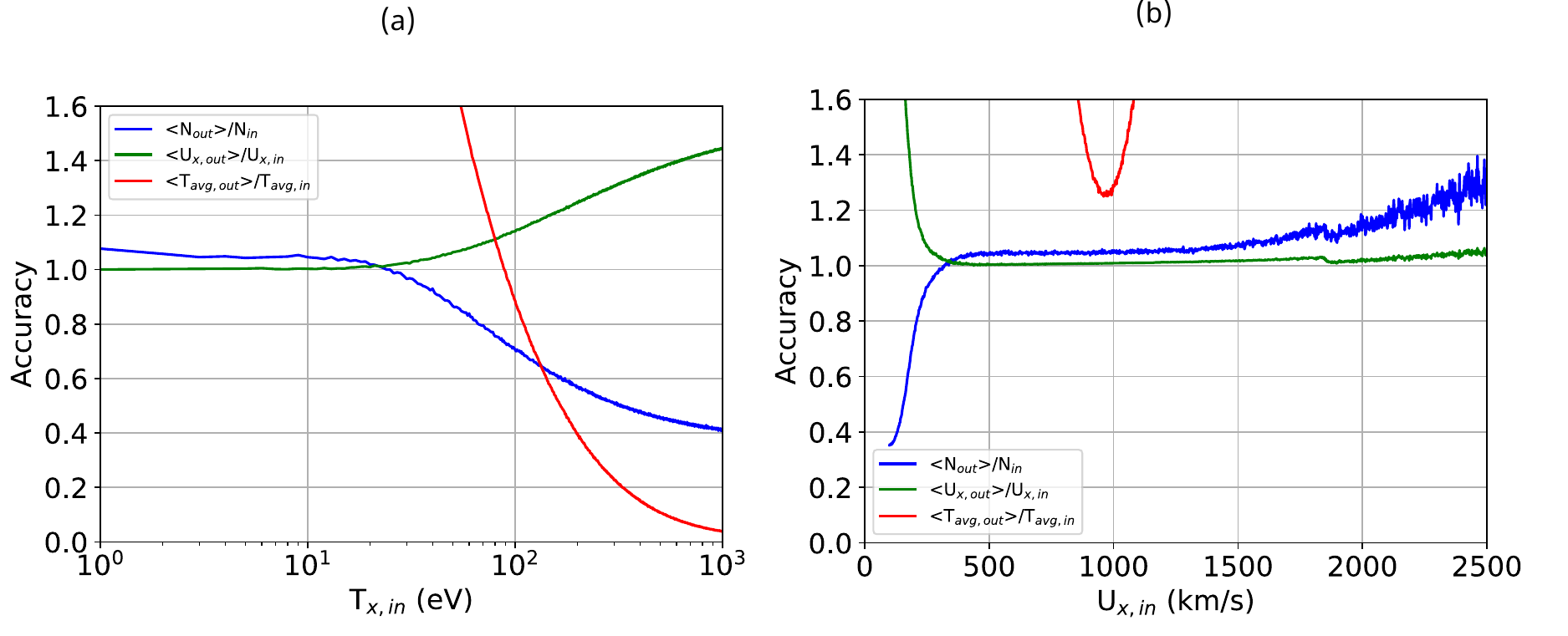}
\caption{
Measurement accuracy for an anisotropic input plasma with a bi-Maxwellian proton core distribution with $N_{\mathrm{in}}=10\,\mathrm{cm}^{-3}$, $U_{y,\mathrm{in}}=U_{z,\mathrm{in}}=0\,\mathrm{km/s}$, and $T_{y,\mathrm{in}}=T_{z,\mathrm{in}}=4T_{x,\mathrm{in}}$, a Mawellian proton beam distribution with $N_{\mathrm{in}}^{\prime}=5\,\mathrm{cm}^{-3}$, $U_{x,\mathrm{in}}^{\prime} = 800\,\mathrm{km/s}$ and ${T_{y,\mathrm{in}}}^{\prime}={T_{z,\mathrm{in}}}^{\prime}={T_{x,\mathrm{in}}}^{\prime}=20\,\mathrm{eV}$, and an added $\alpha$-particle population ($N_{\mathrm{in}}^{\prime\prime}=0.04N_{\mathrm{in}}$, $U_{i,\mathrm{in}}^{\prime\prime}=U_{i,\mathrm{in}}$, and $T_{i,\mathrm{in}}^{\prime\prime}=T_{i,\mathrm{in}}$). Left: Accuracy of  $N_{\mathrm{out}}$ and $U_{x,\mathrm{out}}$ as a function of $T_{x,\mathrm{in}}$ for  $U_{x,\mathrm{in}}=500\,\mathrm{km/s}$. Right: Accuracy of $N_{\mathrm{out}}$ and $U_{x,\mathrm{out}}$ as a function of $U_{x,\mathrm{in}}$ for $T_{x,\mathrm{in}}=10\,\mathrm{eV}$.}
\label{alpha_analysis}
\end{figure}

Figure \ref{alpha_analysis}a shows the accuracy plot of the number density, bulk velocity, and temperature depending on  $T_{\mathrm{in}}$. The measurement accuracy curves at different temperatures are almost the same as without the addition of $\alpha$-particles. 

Figure \ref{alpha_analysis}b shows the accuracy plot for the number density, bulk velocity, and temperature depending on $U_{x,{\mathrm{in}}}$  after adding the proton beam and $\alpha$-particles to the model. The overall trends of the number density and bulk velocity curves in Figure~\ref{alpha_analysis} are similar to the curves in Figure~\ref{proton_beam_analysis}. However,  the number density is slightly overestimated after adding $\alpha$-particles to the model. Moreover, as  $U_{x,\mathrm{in}}$ increases, the measurement accuracies of the number density and of the bulk velocity have a short drop-off at an input velocity of about 1800\,km/s and then continue to rise. 

\section{Discussion and Interpretation}

For completeness, we analyze the impact of a $\kappa$-distribution on the measurements of PLA in~\ref{appendix_kappa}. We find that the presence of non-equilibrium tails in the form of a $\kappa$-distribution do not significantly alter the accuracy of the anticipated PLA measurements. In this section, we discuss the impacts of temperature anisotropy and beams on the measurements. 

\subsection{Proton temperature anisotropy}

As seen in Section~\ref{section:result:Bi-Maxwellian distribution function}, when applying the bi-Maxwellian distribution, the shape of the output VDF  appears stretched in both azimuth and elevation dimensions. This is due to the higher temperatures in the directions that correspond to the directions perpendicular to the magnetic field. At very low parallel temperatures, the measurement accuracy for the number density is actually better than in the Maxwellian situation \cite{Nicolaou_2020} since the spread of counts extends toward larger angular directions in azimuth and elevation in our example, which leads to a better resolution of the larger temperature component. Similarly, large temperatures more easily lead to a greater loss of particles outside the field of view in the direction(s) associated with the larger temperature component(s) (in our case, azimuth and elevation). Therefore, we find that the measurement accuracy of the number density and temperature is worse than for the Maxwellian distribution in hot plasmas \cite{Nicolaou_2020}. For individual temperature components, the accuracies in all directions decrease with increasing input temperature. This effect is stronger in the $y$- and $z$-components than in the $x$-component (i.e., the direction resolved mostly via the narrowly spaced energy bins) which causes the observed underestimation of $T_\perp$.
	
The geometry of the instrument uses spherical coordinates in velocity space. In this geometry, VDFs of constant temperature are more likely to be located within the field of view if the bulk velocity is higher, resulting in more accurate temperature measurements in the angular direction in fast wind. 
We observe that, at high velocities, only the $y$-component of the temperature is underestimated. The reason for this asymmetric behaviour is that our instrument has different resolutions in azimuth and elevation. 

For solar wind with a very low bulk velocity, particles arrive below the lower-energy cutoff of 170 eV. PLA does not detect these particles and thus fails to measure the low-energy part of the full velocity distribution. This effect leads to misestimations of number density, velocity, and temperature.  

The density does not affect the overall shape of the VDF but its variation simply scales the number of counts. This effect can impact the statistical error due to finite counting statistics though when the count level near the center of the VDF is only slightly above the one-count level.

\subsection{Effect of proton beams}

We use a bi-Maxwellian proton core and an isotropic Maxwellian proton beam to model the case of a core-beam plasma in Section~\ref{section:result:Distribution with proton beam}. Since the relative drift between the proton core and proton beam contributes to the second-order moment (i.e, temperature) of the overall distribution, we find a total parallel proton temperature that is greater than the individual temperatures of the core and of the beam alone. 

The accuracy of both number density and bulk speed deteriorate with increasing input temperature. As the input temperature increases, more particles move outside the field of view, so that fewer particles are detected. Due to the normalization in the derivation of the bulk velocity from the first moment of the distribution in Equation~(\ref{PLA_velocity_equation}), this leads to an overestimation of $U_{x,\mathrm{out}}$.

We also examine the effect of different input number densities. Once more, the variation in particle density is not likely to affect the measurement results for the three moments under consideration. We attribute the observed improvement in the accuracy of the density and bulk velocity in Figure~
\ref{proton_beam_analysis}a compared to Figure~\ref{bi-max_analysis}a to the way in which we modify the input temperature. We only modify the core temperature in this plot, so that only part of the core distribution lies outside the field of view at these high temperatures. The contribution of the  beam particles to the total density remains constant though, leading to this apparent improvement in the lower order moments.

We observe a significant deviation between the measured and the  proton-core input temperature after adding proton beams because the measurement of temperature now includes contributions from the individual components and from the relative drift between the components. 
However, the result for the temperature in Table~\ref{proton_beam_table} is consistent with the total proton temperature as defined in Equation~(\ref{total_temp}).

\subsection{Effect of proton beams and $\alpha$-particles}
\label{Discussion:proton beams and alpha particles}

As seen in Section \ref{section:result:Distribution with alpha particles},  $\alpha$-particles appear at greater energies than the proton core in the count maps, although we assume their bulk velocities to be equal ($U_{i,\mathrm{in}}^{\prime\prime }=U_{i,\mathrm{in}}$). Since the electrostatic analyzer cannot distinguish between protons and $\alpha$-particles, we analyse the $\alpha$-particles as if they were protons. Therefore, the $\alpha$-particles appear at higher velocities in the analyzed output VDF. We further find that both the output number density and bulk velocity in the $x$-direction are affected in the presence of  $\alpha$-particles. The magnitude of these changes depends on the relative number density of the $\alpha$-particles.

The trends of our accuracy plots are similar to those without $\alpha$-particles. The low impact of $\alpha$-particles is attributed to the small relative density of the added $\alpha$-particles ($N_{\mathrm{in}}^{\prime\prime}=0.04N_{\mathrm{in}}$ in our calculation), so that the proton input moments still dominate the total output moments. The measurement errors of both number density and bulk velocity gradually rise as the input speed increases. For the same bulk speed of protons and $\alpha$-particles (co-moving species) the $\alpha$-particles have twice the energy-per-charge of protons. Therefore, at a bulk speed above around $1800\,\mathrm{km/s}$, a significant portion of the $\alpha$-particle distribution lies outside the energy-per-charge range of our instrument causing the observed inaccuracies in the plasma parameters. The temperature comparison is not physically meaningful in this case since the temperature of mixed populations is not defined.

\ref{appendix_alphas} presents our results and discussion of the impact of $\alpha$-particles without the presence of a proton beam.

\subsection{Limitations of our model evaluation}
\label{section:limitations}

Our analysis of the  PLA performance is prone to a number of limitations. For example, the actual solar wind conditions are often even more complicated than assumed in our model cases \cite<e.g.,>{Marsch_2006,Verscharen_2019}, so that even the bi-Maxwellian model does not cover the actual shape of the underlying VDF well. 

We only demonstrate the impact of proton temperature anisotropy by assuming the magnetic field is directly pointing to the central look direction of the instrument, which is the $x$-direction in our model. In reality, however, the orientation of the magnetic field and thus of the VDF symmetry axis is arbitrary.  In future studies, it is worthwhile considering different magnetic-field orientations to analyze the geometrical effects of the different resolutions in energy, azimuth, and elevation.

The addition of indistinguishable (in a moment integration) $\alpha$-particles complicates the derivation of the plasma temperature. 
Alternative models, for example, multi-component fit models to the measured distribution would allow us to determine the species moments individually \cite<e.g.,>{Nicolaou_2014}. Alternative approaches include the fitting of the VDF over limited energy ranges \cite<e.g.,>{Nicolaou_2018} or the use of proton beam tracking techniques \cite<e.g.,>{De_2018}. Either way, our analysis clearly leads us to recommend the downlink of count maps for ground moment calculation over the use of onboard moment calculations to avoid inaccuracies in the characterization of the plasma properties. 

\section{Conclusions}

We evaluate the expected performance of the Vigil/PLA instrument under realistic solar wind conditions. Temperature anisotropy impacts the performance of PLA compared to the known behavior for a Maxwellian equilibrium input distribution \cite{Nicolaou_2020}. For example, anisotropic temperatures can deteriorate the accuracy of the measured number density and temperature in high-temperature solar wind compared to the isotropic case. The impact of suprathermal tails in the form of $\kappa$-distributions has a minor impact on the accuracy of the PLA moment determination.

We also study the impact of proton beams and $\alpha$-particles on the measurement accuracy of PLA. We find that adding a proton beam does not affect the number density and velocity measurements much, so that a realistic determination of plasma moments for core-beam plasmas is feasible with PLA. When adding $\alpha$-particles, our instrument treats them as protons, leading to an incorrect distribution function and thus misestimations in the output moment integrations. As expected, these misestimations depend especially on the relative number density of the $\alpha$-particles. The impact on the integration of the supposed proton temperature is particularly strong.

The Vigil/PLA requirements define the required accuracy in all moments as 5\% or better for a number of plasma parameter combinations. Our analysis shows that, depending on the non-equilibrium features in the distribution and the abundance of $\alpha$-particles, deviations of more than 5\% can occur. Ground calculation of the moments, however, would allow for the application of more sophisticated analysis routines (such as fitting) to correct the moments.

Overall, we show that PLA will provide a reliable determination of proton moments within a reasonable range of solar wind parameters, even when the distribution functions are non-equilibrium.  Especially the presence of $\alpha$-particles, however, deteriorates the accuracy of the bulk velocity and the temperature. We, therefore, recommend additional steps to separate the effects of $\alpha$-particles, such as fitting with model distributions
\cite{Nicolaou_2016} or cut-off techniques in velocity space  \cite{Marsch_1982_alpha}. However, these methods require the downlink of full count maps from the Vigil spacecraft, since an automated and unchecked application of these methods on board is unfeasible. For reliable and accurate measurements of the plasma moments, as required from a space-weather monitor asset like Vigil, we therefore recommend the use of ground moments over on-board moments, even if this tradeoff leads to a lower possible measurement cadence due to the limited availability of telemetry bandwidth.

\section{Open Research}

\subsection{Data Available Statement}
All data shown in this study and the Python code that created the figures are publicly available at \url{https://doi.org/10.5281/zenodo.10550337} \cite{dataset23}.

\acknowledgments
This work was supported by STFC Ernest Rutherford Fellowship ST/P003826/1 and STFC Consolidated Grants ST/S000240/1 and ST/W001004/1. We appreciate helpful discussions with the PLA engineering team at MSSL.

\bibliography{PLA_bib}

\begin{thebibliography}{}

\bibitem [\protect \citeauthoryear {%
Alterman%
\ \BBA {} Kasper%
}{%
Alterman%
\ \BBA {} Kasper%
}{%
{\protect \APACyear {2019}}%
}]{%
Alterman_2019}
\APACinsertmetastar {%
Alterman_2019}%
\begin{APACrefauthors}%
Alterman, B\BPBI L.%
\BCBT {}\ \BBA {} Kasper, J\BPBI C.%
\end{APACrefauthors}%
\unskip\
\newblock
\APACrefYearMonthDay{2019}{jun}{}.
\newblock
{\BBOQ}\APACrefatitle {Helium Variation across Two Solar Cycles Reveals a
  Speed-dependent Phase Lag} {Helium variation across two solar cycles reveals
  a speed-dependent phase lag}.{\BBCQ}
\newblock
\APACjournalVolNumPages{The Astrophysical Journal Letters}{879}{1}{L6}.
\newblock
\begin{APACrefURL} \url{https://dx.doi.org/10.3847/2041-8213/ab2391}
  \end{APACrefURL}
\newblock
\begin{APACrefDOI} \doi{10.3847/2041-8213/ab2391} \end{APACrefDOI}
\PrintBackRefs{\CurrentBib}

\bibitem [\protect \citeauthoryear {%
Alterman%
, Kasper%
, Stevens%
\BCBL {}\ \BBA {} Koval%
}{%
Alterman%
\ \protect \BOthers {.}}{%
{\protect \APACyear {2018}}%
}]{%
Alterman_2018}
\APACinsertmetastar {%
Alterman_2018}%
\begin{APACrefauthors}%
Alterman, B\BPBI L.%
, Kasper, J\BPBI C.%
, Stevens, M\BPBI L.%
\BCBL {}\ \BBA {} Koval, A.%
\end{APACrefauthors}%
\unskip\
\newblock
\APACrefYearMonthDay{2018}{sep}{}.
\newblock
{\BBOQ}\APACrefatitle {A Comparison of Alpha Particle and Proton Beam
  Differential Flows in Collisionally Young Solar Wind} {A comparison of alpha
  particle and proton beam differential flows in collisionally young solar
  wind}.{\BBCQ}
\newblock
\APACjournalVolNumPages{The Astrophysical Journal}{864}{2}{112}.
\newblock
\begin{APACrefURL} \url{https://dx.doi.org/10.3847/1538-4357/aad23f}
  \end{APACrefURL}
\newblock
\begin{APACrefDOI} \doi{10.3847/1538-4357/aad23f} \end{APACrefDOI}
\PrintBackRefs{\CurrentBib}

\bibitem [\protect \citeauthoryear {%
Baker%
\ \protect \BOthers {.}}{%
Baker%
\ \protect \BOthers {.}}{%
{\protect \APACyear {2013}}%
}]{%
Baker_2013}
\APACinsertmetastar {%
Baker_2013}%
\begin{APACrefauthors}%
Baker, D\BPBI N.%
, Li, X.%
, Pulkkinen, A.%
, Ngwira, C\BPBI M.%
, Mays, M\BPBI L.%
, Galvin, A\BPBI B.%
\BCBL {}\ \BBA {} Simunac, K\BPBI D\BPBI C.%
\end{APACrefauthors}%
\unskip\
\newblock
\APACrefYearMonthDay{2013}{}{}.
\newblock
{\BBOQ}\APACrefatitle {A major solar eruptive event in July 2012: Defining
  extreme space weather scenarios} {A major solar eruptive event in july 2012:
  Defining extreme space weather scenarios}.{\BBCQ}
\newblock
\APACjournalVolNumPages{Space Weather}{11}{10}{585-591}.
\newblock
\begin{APACrefURL}
  \url{https://agupubs.onlinelibrary.wiley.com/doi/abs/10.1002/swe.20097}
  \end{APACrefURL}
\newblock
\begin{APACrefDOI} \doi{https://doi.org/10.1002/swe.20097} \end{APACrefDOI}
\PrintBackRefs{\CurrentBib}

\bibitem [\protect \citeauthoryear {%
Bale%
\ \protect \BOthers {.}}{%
Bale%
\ \protect \BOthers {.}}{%
{\protect \APACyear {2009}}%
}]{%
Bale_2009}
\APACinsertmetastar {%
Bale_2009}%
\begin{APACrefauthors}%
Bale, S\BPBI D.%
, Kasper, J\BPBI C.%
, Howes, G\BPBI G.%
, Quataert, E.%
, Salem, C.%
\BCBL {}\ \BBA {} Sundkvist, D.%
\end{APACrefauthors}%
\unskip\
\newblock
\APACrefYearMonthDay{2009}{Nov}{}.
\newblock
{\BBOQ}\APACrefatitle {Magnetic Fluctuation Power Near Proton Temperature
  Anisotropy Instability Thresholds in the Solar Wind} {Magnetic fluctuation
  power near proton temperature anisotropy instability thresholds in the solar
  wind}.{\BBCQ}
\newblock
\APACjournalVolNumPages{Phys. Rev. Lett.}{103}{}{211101}.
\newblock
\begin{APACrefURL}
  \url{https://link.aps.org/doi/10.1103/PhysRevLett.103.211101}
  \end{APACrefURL}
\newblock
\begin{APACrefDOI} \doi{10.1103/PhysRevLett.103.211101} \end{APACrefDOI}
\PrintBackRefs{\CurrentBib}

\bibitem [\protect \citeauthoryear {%
Bame%
, Asbridge%
, Feldman%
\BCBL {}\ \BBA {} Gosling%
}{%
Bame%
\ \protect \BOthers {.}}{%
{\protect \APACyear {1977}}%
}]{%
Bame_1977}
\APACinsertmetastar {%
Bame_1977}%
\begin{APACrefauthors}%
Bame, S\BPBI J.%
, Asbridge, J\BPBI R.%
, Feldman, W\BPBI C.%
\BCBL {}\ \BBA {} Gosling, J\BPBI T.%
\end{APACrefauthors}%
\unskip\
\newblock
\APACrefYearMonthDay{1977}{}{}.
\newblock
{\BBOQ}\APACrefatitle {Evidence for a structure-free state at high solar wind
  speeds} {Evidence for a structure-free state at high solar wind
  speeds}.{\BBCQ}
\newblock
\APACjournalVolNumPages{Journal of Geophysical Research
  (1896-1977)}{82}{10}{1487-1492}.
\newblock
\begin{APACrefURL}
  \url{https://agupubs.onlinelibrary.wiley.com/doi/abs/10.1029/JA082i010p01487}
  \end{APACrefURL}
\newblock
\begin{APACrefDOI} \doi{https://doi.org/10.1029/JA082i010p01487}
  \end{APACrefDOI}
\PrintBackRefs{\CurrentBib}

\bibitem [\protect \citeauthoryear {%
Bourouaine%
, Marsch%
\BCBL {}\ \BBA {} Neubauer%
}{%
Bourouaine%
\ \protect \BOthers {.}}{%
{\protect \APACyear {2010}}%
}]{%
Bourouaine_2010}
\APACinsertmetastar {%
Bourouaine_2010}%
\begin{APACrefauthors}%
Bourouaine, S.%
, Marsch, E.%
\BCBL {}\ \BBA {} Neubauer, F\BPBI M.%
\end{APACrefauthors}%
\unskip\
\newblock
\APACrefYearMonthDay{2010}{}{}.
\newblock
{\BBOQ}\APACrefatitle {Correlations between the proton temperature anisotropy
  and transverse high-frequency waves in the solar wind} {Correlations between
  the proton temperature anisotropy and transverse high-frequency waves in the
  solar wind}.{\BBCQ}
\newblock
\APACjournalVolNumPages{Geophysical Research Letters}{37}{14}{}.
\newblock
\begin{APACrefURL}
  \url{https://agupubs.onlinelibrary.wiley.com/doi/abs/10.1029/2010GL043697}
  \end{APACrefURL}
\newblock
\begin{APACrefDOI} \doi{https://doi.org/10.1029/2010GL043697} \end{APACrefDOI}
\PrintBackRefs{\CurrentBib}

\bibitem [\protect \citeauthoryear {%
Cara%
\ \protect \BOthers {.}}{%
Cara%
\ \protect \BOthers {.}}{%
{\protect \APACyear {2017}}%
}]{%
Cara_2017}
\APACinsertmetastar {%
Cara_2017}%
\begin{APACrefauthors}%
Cara, A.%
, Lavraud, B.%
, Fedorov, A.%
, De~Keyser, J.%
, DeMarco, R.%
, Marcucci, M\BPBI F.%
\BDBL {}Bruno, R.%
\end{APACrefauthors}%
\unskip\
\newblock
\APACrefYearMonthDay{2017}{}{}.
\newblock
{\BBOQ}\APACrefatitle {Electrostatic analyzer design for solar wind proton
  measurements with high temporal, energy, and angular resolutions}
  {Electrostatic analyzer design for solar wind proton measurements with high
  temporal, energy, and angular resolutions}.{\BBCQ}
\newblock
\APACjournalVolNumPages{Journal of Geophysical Research: Space
  Physics}{122}{2}{1439-1450}.
\newblock
\begin{APACrefURL}
  \url{https://agupubs.onlinelibrary.wiley.com/doi/abs/10.1002/2016JA023269}
  \end{APACrefURL}
\newblock
\begin{APACrefDOI} \doi{https://doi.org/10.1002/2016JA023269} \end{APACrefDOI}
\PrintBackRefs{\CurrentBib}

\bibitem [\protect \citeauthoryear {%
{Cranmer}%
, {Gibson}%
\BCBL {}\ \BBA {} {Riley}%
}{%
{Cranmer}%
\ \protect \BOthers {.}}{%
{\protect \APACyear {2017}}%
}]{%
Cranmer_2017}
\APACinsertmetastar {%
Cranmer_2017}%
\begin{APACrefauthors}%
{Cranmer}, S\BPBI R.%
, {Gibson}, S\BPBI E.%
\BCBL {}\ \BBA {} {Riley}, P.%
\end{APACrefauthors}%
\unskip\
\newblock
\APACrefYearMonthDay{2017}{{\APACmonth{11}}}{}.
\newblock
{\BBOQ}\APACrefatitle {{Origins of the Ambient Solar Wind: Implications for
  Space Weather}} {{Origins of the Ambient Solar Wind: Implications for Space
  Weather}}.{\BBCQ}
\newblock
\APACjournalVolNumPages{Space Sci.~Rev.}{212}{3-4}{1345-1384}.
\newblock
\begin{APACrefDOI} \doi{10.1007/s11214-017-0416-y} \end{APACrefDOI}
\PrintBackRefs{\CurrentBib}

\bibitem [\protect \citeauthoryear {%
De~Keyser%
\ \protect \BOthers {.}}{%
De~Keyser%
\ \protect \BOthers {.}}{%
{\protect \APACyear {2018}}%
}]{%
De_2018}
\APACinsertmetastar {%
De_2018}%
\begin{APACrefauthors}%
De~Keyser, J.%
, Lavraud, B.%
, P\v{r}ech, L.%
, Neefs, E.%
, Berkenbosch, S.%
, Beeckman, B.%
\BDBL {}Brienza, D.%
\end{APACrefauthors}%
\unskip\
\newblock
\APACrefYearMonthDay{2018}{}{}.
\newblock
{\BBOQ}\APACrefatitle {Beam tracking strategies for fast acquisition of solar
  wind velocity distribution functions with high energy and angular
  resolutions} {Beam tracking strategies for fast acquisition of solar wind
  velocity distribution functions with high energy and angular
  resolutions}.{\BBCQ}
\newblock
\APACjournalVolNumPages{Annales Geophysicae}{36}{5}{1285--1302}.
\newblock
\begin{APACrefURL} \url{https://angeo.copernicus.org/articles/36/1285/2018/}
  \end{APACrefURL}
\newblock
\begin{APACrefDOI} \doi{10.5194/angeo-36-1285-2018} \end{APACrefDOI}
\PrintBackRefs{\CurrentBib}

\bibitem [\protect \citeauthoryear {%
de Koning%
\ \BBA {} Pizzo%
}{%
de Koning%
\ \BBA {} Pizzo%
}{%
{\protect \APACyear {2011}}%
}]{%
dekoning_2011}
\APACinsertmetastar {%
dekoning_2011}%
\begin{APACrefauthors}%
de Koning, C\BPBI A.%
\BCBT {}\ \BBA {} Pizzo, V\BPBI J.%
\end{APACrefauthors}%
\unskip\
\newblock
\APACrefYearMonthDay{2011}{}{}.
\newblock
{\BBOQ}\APACrefatitle {Polarimetric localization: A new tool for calculating
  the CME speed and direction of propagation in near-real time} {Polarimetric
  localization: A new tool for calculating the cme speed and direction of
  propagation in near-real time}.{\BBCQ}
\newblock
\APACjournalVolNumPages{Space Weather}{9}{3}{}.
\newblock
\begin{APACrefURL}
  \url{https://agupubs.onlinelibrary.wiley.com/doi/abs/10.1029/2010SW000595}
  \end{APACrefURL}
\newblock
\begin{APACrefDOI} \doi{https://doi.org/10.1029/2010SW000595} \end{APACrefDOI}
\PrintBackRefs{\CurrentBib}

\bibitem [\protect \citeauthoryear {%
Eastwood%
\ \protect \BOthers {.}}{%
Eastwood%
\ \protect \BOthers {.}}{%
{\protect \APACyear {2017}}%
}]{%
Eastwood_2017}
\APACinsertmetastar {%
Eastwood_2017}%
\begin{APACrefauthors}%
Eastwood, J\BPBI P.%
, Biffis, E.%
, Hapgood, M\BPBI A.%
, Green, L.%
, Bisi, M\BPBI M.%
, Bentley, R\BPBI D.%
\BDBL {}Burnett, C.%
\end{APACrefauthors}%
\unskip\
\newblock
\APACrefYearMonthDay{2017}{}{}.
\newblock
{\BBOQ}\APACrefatitle {The Economic Impact of Space Weather: Where Do We
  Stand?} {The economic impact of space weather: Where do we stand?}{\BBCQ}
\newblock
\APACjournalVolNumPages{Risk Analysis}{37}{2}{206-218}.
\newblock
\begin{APACrefURL}
  \url{https://onlinelibrary.wiley.com/doi/abs/10.1111/risa.12765}
  \end{APACrefURL}
\newblock
\begin{APACrefDOI} \doi{https://doi.org/10.1111/risa.12765} \end{APACrefDOI}
\PrintBackRefs{\CurrentBib}

\bibitem [\protect \citeauthoryear {%
{Eastwood}%
, {Nakamura}%
, {Turc}%
, {Mejnertsen}%
\BCBL {}\ \BBA {} {Hesse}%
}{%
{Eastwood}%
\ \protect \BOthers {.}}{%
{\protect \APACyear {2017}}%
}]{%
Eastwood_2017b}
\APACinsertmetastar {%
Eastwood_2017b}%
\begin{APACrefauthors}%
{Eastwood}, J\BPBI P.%
, {Nakamura}, R.%
, {Turc}, L.%
, {Mejnertsen}, L.%
\BCBL {}\ \BBA {} {Hesse}, M.%
\end{APACrefauthors}%
\unskip\
\newblock
\APACrefYearMonthDay{2017}{{\APACmonth{11}}}{}.
\newblock
{\BBOQ}\APACrefatitle {{The Scientific Foundations of Forecasting
  Magnetospheric Space Weather}} {{The Scientific Foundations of Forecasting
  Magnetospheric Space Weather}}.{\BBCQ}
\newblock
\APACjournalVolNumPages{Space Sci.~Rev.}{212}{3-4}{1221-1252}.
\newblock
\begin{APACrefDOI} \doi{10.1007/s11214-017-0399-8} \end{APACrefDOI}
\PrintBackRefs{\CurrentBib}

\bibitem [\protect \citeauthoryear {%
Hapgood%
}{%
Hapgood%
}{%
{\protect \APACyear {2011}}%
}]{%
Hapgood_2011}
\APACinsertmetastar {%
Hapgood_2011}%
\begin{APACrefauthors}%
Hapgood, M.%
\end{APACrefauthors}%
\unskip\
\newblock
\APACrefYearMonthDay{2011}{}{}.
\newblock
{\BBOQ}\APACrefatitle {Towards a scientific understanding of the risk from
  extreme space weather} {Towards a scientific understanding of the risk from
  extreme space weather}.{\BBCQ}
\newblock
\APACjournalVolNumPages{Advances in Space Research}{47}{12}{2059-2072}.
\newblock
\begin{APACrefURL}
  \url{https://www.sciencedirect.com/science/article/pii/S0273117710001122}
  \end{APACrefURL}
\newblock
\APACrefnote{Recent Advances in Space Weather Monitoring, Modelling, and
  Forecasting - 2}
\newblock
\begin{APACrefDOI} \doi{https://doi.org/10.1016/j.asr.2010.02.007}
  \end{APACrefDOI}
\PrintBackRefs{\CurrentBib}

\bibitem [\protect \citeauthoryear {%
Hellinger%
, Trávníček%
, Kasper%
\BCBL {}\ \BBA {} Lazarus%
}{%
Hellinger%
\ \protect \BOthers {.}}{%
{\protect \APACyear {2006}}%
}]{%
Hellinger_2006}
\APACinsertmetastar {%
Hellinger_2006}%
\begin{APACrefauthors}%
Hellinger, P.%
, Trávníček, P.%
, Kasper, J\BPBI C.%
\BCBL {}\ \BBA {} Lazarus, A\BPBI J.%
\end{APACrefauthors}%
\unskip\
\newblock
\APACrefYearMonthDay{2006}{}{}.
\newblock
{\BBOQ}\APACrefatitle {Solar wind proton temperature anisotropy: Linear theory
  and WIND/SWE observations} {Solar wind proton temperature anisotropy: Linear
  theory and wind/swe observations}.{\BBCQ}
\newblock
\APACjournalVolNumPages{Geophysical Research Letters}{33}{9}{}.
\newblock
\begin{APACrefURL}
  \url{https://agupubs.onlinelibrary.wiley.com/doi/abs/10.1029/2006GL025925}
  \end{APACrefURL}
\newblock
\begin{APACrefDOI} \doi{https://doi.org/10.1029/2006GL025925} \end{APACrefDOI}
\PrintBackRefs{\CurrentBib}

\bibitem [\protect \citeauthoryear {%
Kasper%
, Lazarus%
, Steinberg%
, Ogilvie%
\BCBL {}\ \BBA {} Szabo%
}{%
Kasper%
\ \protect \BOthers {.}}{%
{\protect \APACyear {2006}}%
}]{%
kasper_2006}
\APACinsertmetastar {%
kasper_2006}%
\begin{APACrefauthors}%
Kasper, J\BPBI C.%
, Lazarus, A\BPBI J.%
, Steinberg, J\BPBI T.%
, Ogilvie, K\BPBI W.%
\BCBL {}\ \BBA {} Szabo, A.%
\end{APACrefauthors}%
\unskip\
\newblock
\APACrefYearMonthDay{2006}{}{}.
\newblock
{\BBOQ}\APACrefatitle {Physics-based tests to identify the accuracy of solar
  wind ion measurements: A case study with the Wind Faraday Cups}
  {Physics-based tests to identify the accuracy of solar wind ion measurements:
  A case study with the wind faraday cups}.{\BBCQ}
\newblock
\APACjournalVolNumPages{Journal of Geophysical Research: Space
  Physics}{111}{A3}{}.
\newblock
\begin{APACrefURL}
  \url{https://agupubs.onlinelibrary.wiley.com/doi/abs/10.1029/2005JA011442}
  \end{APACrefURL}
\newblock
\begin{APACrefDOI} \doi{https://doi.org/10.1029/2005JA011442} \end{APACrefDOI}
\PrintBackRefs{\CurrentBib}

\bibitem [\protect \citeauthoryear {%
Knoll%
\ \BBA {} Glenn%
}{%
Knoll%
\ \BBA {} Glenn%
}{%
{\protect \APACyear {1989}}%
}]{%
Knoll_1989}
\APACinsertmetastar {%
Knoll_1989}%
\begin{APACrefauthors}%
Knoll%
\BCBT {}\ \BBA {} Glenn, F.%
\end{APACrefauthors}%
\unskip\
\newblock
\APACrefYear{1989}.
\newblock
\APACrefbtitle {Radiation detection and measurement / Glenn F. Knoll}
  {Radiation detection and measurement / glenn f. knoll}.
\newblock
\APACaddressPublisher{}{Radiation detection and measurement}.
\PrintBackRefs{\CurrentBib}

\bibitem [\protect \citeauthoryear {%
Liou%
\ \protect \BOthers {.}}{%
Liou%
\ \protect \BOthers {.}}{%
{\protect \APACyear {2014}}%
}]{%
LiouKan_2014}
\APACinsertmetastar {%
LiouKan_2014}%
\begin{APACrefauthors}%
Liou, K.%
, Wu, C\BHBI C.%
, Dryer, M.%
, Wu, S\BHBI T.%
, Rich, N.%
, Plunkett, S.%
\BDBL {}Schenk, K.%
\end{APACrefauthors}%
\unskip\
\newblock
\APACrefYearMonthDay{2014}{}{}.
\newblock
{\BBOQ}\APACrefatitle {Global simulation of extremely fast coronal mass
  ejection on 23 July 2012} {Global simulation of extremely fast coronal mass
  ejection on 23 july 2012}.{\BBCQ}
\newblock
\APACjournalVolNumPages{Journal of atmospheric and solar-terrestrial
  physics}{121}{}{32-41}.
\PrintBackRefs{\CurrentBib}

\bibitem [\protect \citeauthoryear {%
Livadiotis%
\ \BBA {} McComas%
}{%
Livadiotis%
\ \BBA {} McComas%
}{%
{\protect \APACyear {2009}}%
}]{%
Livadiotis_2009}
\APACinsertmetastar {%
Livadiotis_2009}%
\begin{APACrefauthors}%
Livadiotis, G.%
\BCBT {}\ \BBA {} McComas, D\BPBI J.%
\end{APACrefauthors}%
\unskip\
\newblock
\APACrefYearMonthDay{2009}{}{}.
\newblock
{\BBOQ}\APACrefatitle {Beyond kappa distributions: Exploiting Tsallis
  statistical mechanics in space plasmas} {Beyond kappa distributions:
  Exploiting tsallis statistical mechanics in space plasmas}.{\BBCQ}
\newblock
\APACjournalVolNumPages{Journal of Geophysical Research: Space
  Physics}{114}{A11}{}.
\newblock
\begin{APACrefURL}
  \url{https://agupubs.onlinelibrary.wiley.com/doi/abs/10.1029/2009JA014352}
  \end{APACrefURL}
\newblock
\begin{APACrefDOI} \doi{https://doi.org/10.1029/2009JA014352} \end{APACrefDOI}
\PrintBackRefs{\CurrentBib}

\bibitem [\protect \citeauthoryear {%
Livadiotis%
\ \BBA {} McComas%
}{%
Livadiotis%
\ \BBA {} McComas%
}{%
{\protect \APACyear {2013}}%
}]{%
Livadiotis_2013}
\APACinsertmetastar {%
Livadiotis_2013}%
\begin{APACrefauthors}%
Livadiotis, G.%
\BCBT {}\ \BBA {} McComas, D\BPBI J.%
\end{APACrefauthors}%
\unskip\
\newblock
\APACrefYearMonthDay{2013}{Jun}{01}.
\newblock
{\BBOQ}\APACrefatitle {Understanding Kappa Distributions: A Toolbox for Space
  Science and Astrophysics} {Understanding kappa distributions: A toolbox for
  space science and astrophysics}.{\BBCQ}
\newblock
\APACjournalVolNumPages{Space Science Reviews}{175}{1}{183-214}.
\newblock
\begin{APACrefURL} \url{https://doi.org/10.1007/s11214-013-9982-9}
  \end{APACrefURL}
\newblock
\begin{APACrefDOI} \doi{10.1007/s11214-013-9982-9} \end{APACrefDOI}
\PrintBackRefs{\CurrentBib}

\bibitem [\protect \citeauthoryear {%
Marsch%
}{%
Marsch%
}{%
{\protect \APACyear {2006}}%
}]{%
Marsch_2006}
\APACinsertmetastar {%
Marsch_2006}%
\begin{APACrefauthors}%
Marsch, E.%
\end{APACrefauthors}%
\unskip\
\newblock
\APACrefYearMonthDay{2006}{Jul}{27}.
\newblock
{\BBOQ}\APACrefatitle {Kinetic Physics of the Solar Corona and Solar Wind}
  {Kinetic physics of the solar corona and solar wind}.{\BBCQ}
\newblock
\APACjournalVolNumPages{Living Reviews in Solar Physics}{3}{1}{1}.
\newblock
\begin{APACrefURL} \url{https://doi.org/10.12942/lrsp-2006-1} \end{APACrefURL}
\newblock
\begin{APACrefDOI} \doi{10.12942/lrsp-2006-1} \end{APACrefDOI}
\PrintBackRefs{\CurrentBib}

\bibitem [\protect \citeauthoryear {%
Marsch%
, Ao%
\BCBL {}\ \BBA {} Tu%
}{%
Marsch%
\ \protect \BOthers {.}}{%
{\protect \APACyear {2004}}%
}]{%
Marsch_2004}
\APACinsertmetastar {%
Marsch_2004}%
\begin{APACrefauthors}%
Marsch, E.%
, Ao, X\BHBI Z.%
\BCBL {}\ \BBA {} Tu, C\BHBI Y.%
\end{APACrefauthors}%
\unskip\
\newblock
\APACrefYearMonthDay{2004}{}{}.
\newblock
{\BBOQ}\APACrefatitle {On the temperature anisotropy of the core part of the
  proton velocity distribution function in the solar wind} {On the temperature
  anisotropy of the core part of the proton velocity distribution function in
  the solar wind}.{\BBCQ}
\newblock
\APACjournalVolNumPages{Journal of Geophysical Research: Space
  Physics}{109}{A4}{}.
\newblock
\begin{APACrefURL}
  \url{https://agupubs.onlinelibrary.wiley.com/doi/abs/10.1029/2003JA010330}
  \end{APACrefURL}
\newblock
\begin{APACrefDOI} \doi{https://doi.org/10.1029/2003JA010330} \end{APACrefDOI}
\PrintBackRefs{\CurrentBib}

\bibitem [\protect \citeauthoryear {%
Marsch%
, Mühlhäuser%
, Rosenbauer%
, Schwenn%
\BCBL {}\ \BBA {} Denskat%
}{%
Marsch%
\ \protect \BOthers {.}}{%
{\protect \APACyear {1981}}%
}]{%
Marsch_1981}
\APACinsertmetastar {%
Marsch_1981}%
\begin{APACrefauthors}%
Marsch, E.%
, Mühlhäuser, K\BHBI H.%
, Rosenbauer, H.%
, Schwenn, R.%
\BCBL {}\ \BBA {} Denskat, K\BPBI U.%
\end{APACrefauthors}%
\unskip\
\newblock
\APACrefYearMonthDay{1981}{}{}.
\newblock
{\BBOQ}\APACrefatitle {Pronounced proton core temperature anisotropy, ion
  differential speed, and simultaneous Alfvén wave activity in slow solar wind
  at 0.3 AU} {Pronounced proton core temperature anisotropy, ion differential
  speed, and simultaneous alfvén wave activity in slow solar wind at 0.3
  au}.{\BBCQ}
\newblock
\APACjournalVolNumPages{Journal of Geophysical Research: Space
  Physics}{86}{A11}{9199-9203}.
\newblock
\begin{APACrefURL}
  \url{https://agupubs.onlinelibrary.wiley.com/doi/abs/10.1029/JA086iA11p09199}
  \end{APACrefURL}
\newblock
\begin{APACrefDOI} \doi{https://doi.org/10.1029/JA086iA11p09199}
  \end{APACrefDOI}
\PrintBackRefs{\CurrentBib}

\bibitem [\protect \citeauthoryear {%
Marsch%
\ \BBA {} Richter%
}{%
Marsch%
\ \BBA {} Richter%
}{%
{\protect \APACyear {1984}}%
}]{%
Marsch_1984}
\APACinsertmetastar {%
Marsch_1984}%
\begin{APACrefauthors}%
Marsch, E.%
\BCBT {}\ \BBA {} Richter, A\BPBI K.%
\end{APACrefauthors}%
\unskip\
\newblock
\APACrefYearMonthDay{1984}{}{}.
\newblock
{\BBOQ}\APACrefatitle {Distribution of solar wind angular momentum between
  particles and magnetic field: Inferences about the Alfvén critical point
  from Helios observations} {Distribution of solar wind angular momentum
  between particles and magnetic field: Inferences about the alfvén critical
  point from helios observations}.{\BBCQ}
\newblock
\APACjournalVolNumPages{Journal of Geophysical Research: Space
  Physics}{89}{A7}{5386-5394}.
\newblock
\begin{APACrefURL}
  \url{https://agupubs.onlinelibrary.wiley.com/doi/abs/10.1029/JA089iA07p05386}
  \end{APACrefURL}
\newblock
\begin{APACrefDOI} \doi{https://doi.org/10.1029/JA089iA07p05386}
  \end{APACrefDOI}
\PrintBackRefs{\CurrentBib}

\bibitem [\protect \citeauthoryear {%
{Marsch}%
, {Rosenbauer}%
, {Schwenn}%
, {Muehlhaeuser}%
\BCBL {}\ \BBA {} {Neubauer}%
}{%
{Marsch}%
, {Rosenbauer}%
\BCBL {}\ \protect \BOthers {.}}{%
{\protect \APACyear {1982}}%
}]{%
Marsch_1982_alpha}
\APACinsertmetastar {%
Marsch_1982_alpha}%
\begin{APACrefauthors}%
{Marsch}, E.%
, {Rosenbauer}, H.%
, {Schwenn}, R.%
, {Muehlhaeuser}, K\BPBI H.%
\BCBL {}\ \BBA {} {Neubauer}, F\BPBI M.%
\end{APACrefauthors}%
\unskip\
\newblock
\APACrefYearMonthDay{1982}{{\APACmonth{01}}}{}.
\newblock
{\BBOQ}\APACrefatitle {{Solar wind helium ions: obsevations of the Helios solar
  probes between 0.3 and 1 AU}} {{Solar wind helium ions: obsevations of the
  Helios solar probes between 0.3 and 1 AU}}.{\BBCQ}
\newblock
\APACjournalVolNumPages{J.~Geophys.~Res.}{87}{A1}{35-51}.
\newblock
\begin{APACrefDOI} \doi{10.1029/JA087iA01p00035} \end{APACrefDOI}
\PrintBackRefs{\CurrentBib}

\bibitem [\protect \citeauthoryear {%
{Marsch}%
, {Schwenn}%
\BCBL {}\ \protect \BOthers {.}}{%
{Marsch}%
, {Schwenn}%
\BCBL {}\ \protect \BOthers {.}}{%
{\protect \APACyear {1982}}%
}]{%
Marsch_1982_proton}
\APACinsertmetastar {%
Marsch_1982_proton}%
\begin{APACrefauthors}%
{Marsch}, E.%
, {Schwenn}, R.%
, {Rosenbauer}, H.%
, {Muehlhaeuser}, K\BPBI H.%
, {Pilipp}, W.%
\BCBL {}\ \BBA {} {Neubauer}, F\BPBI M.%
\end{APACrefauthors}%
\unskip\
\newblock
\APACrefYearMonthDay{1982}{{\APACmonth{01}}}{}.
\newblock
{\BBOQ}\APACrefatitle {{Solar wind protons: Three-dimensional velocity
  distributions and derived plasma parameters measured between 0.3 and 1 AU}}
  {{Solar wind protons: Three-dimensional velocity distributions and derived
  plasma parameters measured between 0.3 and 1 AU}}.{\BBCQ}
\newblock
\APACjournalVolNumPages{J.~Geophys.~Res.}{87}{A1}{52-72}.
\newblock
\begin{APACrefDOI} \doi{10.1029/JA087iA01p00052} \end{APACrefDOI}
\PrintBackRefs{\CurrentBib}

\bibitem [\protect \citeauthoryear {%
Mishra%
\ \BBA {} Srivastava%
}{%
Mishra%
\ \BBA {} Srivastava%
}{%
{\protect \APACyear {2013}}%
}]{%
Mishra_2013}
\APACinsertmetastar {%
Mishra_2013}%
\begin{APACrefauthors}%
Mishra, W.%
\BCBT {}\ \BBA {} Srivastava, N.%
\end{APACrefauthors}%
\unskip\
\newblock
\APACrefYearMonthDay{2013}{jul}{}.
\newblock
{\BBOQ}\APACrefatitle {ESTIMATING THE ARRIVAL TIME OF EARTH-DIRECTED CORONAL
  MASS EJECTIONS AT IN SITU SPACECRAFT USING COR AND HI OBSERVATIONS FROM
  STEREO} {Estimating the arrival time of earth-directed coronal mass ejections
  at in situ spacecraft using cor and hi observations from stereo}.{\BBCQ}
\newblock
\APACjournalVolNumPages{The Astrophysical Journal}{772}{1}{70}.
\newblock
\begin{APACrefURL} \url{https://dx.doi.org/10.1088/0004-637X/772/1/70}
  \end{APACrefURL}
\newblock
\begin{APACrefDOI} \doi{10.1088/0004-637X/772/1/70} \end{APACrefDOI}
\PrintBackRefs{\CurrentBib}

\bibitem [\protect \citeauthoryear {%
Nicolaou%
, Haythornthwaite%
\BCBL {}\ \BBA {} Coates%
}{%
Nicolaou%
\ \protect \BOthers {.}}{%
{\protect \APACyear {2022}}%
}]{%
Nicolaou_2022}
\APACinsertmetastar {%
Nicolaou_2022}%
\begin{APACrefauthors}%
Nicolaou, G.%
, Haythornthwaite, R\BPBI P.%
\BCBL {}\ \BBA {} Coates, A\BPBI J.%
\end{APACrefauthors}%
\unskip\
\newblock
\APACrefYearMonthDay{2022}{}{}.
\newblock
{\BBOQ}\APACrefatitle {Resolving Space Plasma Species With Electrostatic
  Analyzers} {Resolving space plasma species with electrostatic
  analyzers}.{\BBCQ}
\newblock
\APACjournalVolNumPages{Frontiers in Astronomy and Space Sciences}{9}{}{}.
\newblock
\begin{APACrefURL}
  \url{https://www.frontiersin.org/articles/10.3389/fspas.2022.861433)}
  \end{APACrefURL}
\newblock
\begin{APACrefDOI} \doi{10.3389/fspas.2022.861433} \end{APACrefDOI}
\PrintBackRefs{\CurrentBib}

\bibitem [\protect \citeauthoryear {%
Nicolaou%
\ \BBA {} Livadiotis%
}{%
Nicolaou%
\ \BBA {} Livadiotis%
}{%
{\protect \APACyear {2016}}%
}]{%
Nicolaou_2016}
\APACinsertmetastar {%
Nicolaou_2016}%
\begin{APACrefauthors}%
Nicolaou, G.%
\BCBT {}\ \BBA {} Livadiotis, G.%
\end{APACrefauthors}%
\unskip\
\newblock
\APACrefYearMonthDay{2016}{Oct}{13}.
\newblock
{\BBOQ}\APACrefatitle {Misestimation of temperature when applying Maxwellian
  distributions to space plasmas described by kappa distributions}
  {Misestimation of temperature when applying maxwellian distributions to space
  plasmas described by kappa distributions}.{\BBCQ}
\newblock
\APACjournalVolNumPages{Astrophysics and Space Science}{361}{11}{359}.
\newblock
\begin{APACrefURL} \url{https://doi.org/10.1007/s10509-016-2949-z}
  \end{APACrefURL}
\newblock
\begin{APACrefDOI} \doi{10.1007/s10509-016-2949-z} \end{APACrefDOI}
\PrintBackRefs{\CurrentBib}

\bibitem [\protect \citeauthoryear {%
Nicolaou%
, Livadiotis%
, Owen%
, Verscharen%
\BCBL {}\ \BBA {} Wicks%
}{%
Nicolaou%
\ \protect \BOthers {.}}{%
{\protect \APACyear {2018}}%
}]{%
Nicolaou_2018}
\APACinsertmetastar {%
Nicolaou_2018}%
\begin{APACrefauthors}%
Nicolaou, G.%
, Livadiotis, G.%
, Owen, C\BPBI J.%
, Verscharen, D.%
\BCBL {}\ \BBA {} Wicks, R\BPBI T.%
\end{APACrefauthors}%
\unskip\
\newblock
\APACrefYearMonthDay{2018}{aug}{}.
\newblock
{\BBOQ}\APACrefatitle {Determining the Kappa Distributions of Space Plasmas
  from Observations in a Limited Energy Range} {Determining the kappa
  distributions of space plasmas from observations in a limited energy
  range}.{\BBCQ}
\newblock
\APACjournalVolNumPages{The Astrophysical Journal}{864}{1}{3}.
\newblock
\begin{APACrefURL} \url{https://dx.doi.org/10.3847/1538-4357/aad45d}
  \end{APACrefURL}
\newblock
\begin{APACrefDOI} \doi{10.3847/1538-4357/aad45d} \end{APACrefDOI}
\PrintBackRefs{\CurrentBib}

\bibitem [\protect \citeauthoryear {%
Nicolaou%
, McComas%
, Bagenal%
\BCBL {}\ \BBA {} Elliott%
}{%
Nicolaou%
\ \protect \BOthers {.}}{%
{\protect \APACyear {2014}}%
}]{%
Nicolaou_2014}
\APACinsertmetastar {%
Nicolaou_2014}%
\begin{APACrefauthors}%
Nicolaou, G.%
, McComas, D\BPBI J.%
, Bagenal, F.%
\BCBL {}\ \BBA {} Elliott, H\BPBI A.%
\end{APACrefauthors}%
\unskip\
\newblock
\APACrefYearMonthDay{2014}{}{}.
\newblock
{\BBOQ}\APACrefatitle {Properties of plasma ions in the distant Jovian
  magnetosheath using Solar Wind Around Pluto data on New Horizons} {Properties
  of plasma ions in the distant jovian magnetosheath using solar wind around
  pluto data on new horizons}.{\BBCQ}
\newblock
\APACjournalVolNumPages{Journal of Geophysical Research: Space
  Physics}{119}{5}{3463-3479}.
\newblock
\begin{APACrefURL}
  \url{https://agupubs.onlinelibrary.wiley.com/doi/abs/10.1002/2013JA019665}
  \end{APACrefURL}
\newblock
\begin{APACrefDOI} \doi{https://doi.org/10.1002/2013JA019665} \end{APACrefDOI}
\PrintBackRefs{\CurrentBib}

\bibitem [\protect \citeauthoryear {%
Nicolaou%
, Wicks%
, Rae%
\BCBL {}\ \BBA {} Kataria%
}{%
Nicolaou%
\ \protect \BOthers {.}}{%
{\protect \APACyear {2020}}%
}]{%
Nicolaou_2020}
\APACinsertmetastar {%
Nicolaou_2020}%
\begin{APACrefauthors}%
Nicolaou, G.%
, Wicks, R\BPBI T.%
, Rae, I\BPBI J.%
\BCBL {}\ \BBA {} Kataria, D\BPBI O.%
\end{APACrefauthors}%
\unskip\
\newblock
\APACrefYearMonthDay{2020}{}{}.
\newblock
{\BBOQ}\APACrefatitle {Evaluating the Performance of a Plasma Analyzer for a
  Space Weather Monitor Mission Concept} {Evaluating the performance of a
  plasma analyzer for a space weather monitor mission concept}.{\BBCQ}
\newblock
\APACjournalVolNumPages{Space Weather}{18}{12}{e2020SW002559}.
\newblock
\begin{APACrefURL}
  \url{https://agupubs.onlinelibrary.wiley.com/doi/abs/10.1029/2020SW002559}
  \end{APACrefURL}
\newblock
\APACrefnote{e2020SW002559 10.1029/2020SW002559}
\newblock
\begin{APACrefDOI} \doi{https://doi.org/10.1029/2020SW002559} \end{APACrefDOI}
\PrintBackRefs{\CurrentBib}

\bibitem [\protect \citeauthoryear {%
Oughton%
\ \protect \BOthers {.}}{%
Oughton%
\ \protect \BOthers {.}}{%
{\protect \APACyear {2016}}%
}]{%
Oughton_2016}
\APACinsertmetastar {%
Oughton_2016}%
\begin{APACrefauthors}%
Oughton, E.%
, Copic, J.%
, Skelton, A.%
, Kesaite, V.%
, Yeo, Z\BPBI Y.%
, Ruffle, S\BPBI J.%
\BDBL {}Ralph, D.%
\end{APACrefauthors}%
\unskip\
\newblock
\APACrefYearMonthDay{2016}{}{}.
\newblock
\APACrefbtitle {Helios Solar Storm Scenario} {Helios solar storm scenario}\
  \APACbVolEdTR{}{\BTR{}}.
\newblock
\APACaddressInstitution{Centre for Risk Studies}{University of Cambridge}.
\PrintBackRefs{\CurrentBib}

\bibitem [\protect \citeauthoryear {%
{Owen}%
\ \protect \BOthers {.}}{%
{Owen}%
\ \protect \BOthers {.}}{%
{\protect \APACyear {2020}}%
}]{%
Owen_2020}
\APACinsertmetastar {%
Owen_2020}%
\begin{APACrefauthors}%
{Owen}, C\BPBI J.%
, {Bruno}, R.%
, {Livi}, S.%
, {Louarn}, P.%
, {Al Janabi}, K.%
, {Allegrini}, F.%
\BDBL {}{Zouganelis}, I.%
\end{APACrefauthors}%
\unskip\
\newblock
\APACrefYearMonthDay{2020}{{\APACmonth{10}}}{}.
\newblock
{\BBOQ}\APACrefatitle {{The Solar Orbiter Solar Wind Analyser (SWA) suite}}
  {{The Solar Orbiter Solar Wind Analyser (SWA) suite}}.{\BBCQ}
\newblock
\APACjournalVolNumPages{Astron.~Astrophys.}{642}{}{A16}.
\newblock
\begin{APACrefDOI} \doi{10.1051/0004-6361/201937259} \end{APACrefDOI}
\PrintBackRefs{\CurrentBib}

\bibitem [\protect \citeauthoryear {%
Schrijver%
\ \protect \BOthers {.}}{%
Schrijver%
\ \protect \BOthers {.}}{%
{\protect \APACyear {2015}}%
}]{%
SCHRIJVER_2015}
\APACinsertmetastar {%
SCHRIJVER_2015}%
\begin{APACrefauthors}%
Schrijver, C\BPBI J.%
, Kauristie, K.%
, Aylward, A\BPBI D.%
, Denardini, C\BPBI M.%
, Gibson, S\BPBI E.%
, Glover, A.%
\BDBL {}Vilmer, N.%
\end{APACrefauthors}%
\unskip\
\newblock
\APACrefYearMonthDay{2015}{}{}.
\newblock
{\BBOQ}\APACrefatitle {Understanding space weather to shield society: A global
  road map for 2015–2025 commissioned by COSPAR and ILWS} {Understanding
  space weather to shield society: A global road map for 2015–2025
  commissioned by cospar and ilws}.{\BBCQ}
\newblock
\APACjournalVolNumPages{Advances in Space Research}{55}{12}{2745-2807}.
\newblock
\begin{APACrefURL}
  \url{https://www.sciencedirect.com/science/article/pii/S0273117715002252}
  \end{APACrefURL}
\newblock
\begin{APACrefDOI} \doi{https://doi.org/10.1016/j.asr.2015.03.023}
  \end{APACrefDOI}
\PrintBackRefs{\CurrentBib}

\bibitem [\protect \citeauthoryear {%
{Schwenn}%
}{%
{Schwenn}%
}{%
{\protect \APACyear {2006}}%
}]{%
Schwenn_2006}
\APACinsertmetastar {%
Schwenn_2006}%
\begin{APACrefauthors}%
{Schwenn}, R.%
\end{APACrefauthors}%
\unskip\
\newblock
\APACrefYearMonthDay{2006}{{\APACmonth{08}}}{}.
\newblock
{\BBOQ}\APACrefatitle {{Space Weather: The Solar Perspective}} {{Space Weather:
  The Solar Perspective}}.{\BBCQ}
\newblock
\APACjournalVolNumPages{Living Reviews in Solar Physics}{3}{1}{2}.
\newblock
\begin{APACrefDOI} \doi{10.12942/lrsp-2006-2} \end{APACrefDOI}
\PrintBackRefs{\CurrentBib}

\bibitem [\protect \citeauthoryear {%
Stone%
\ \protect \BOthers {.}}{%
Stone%
\ \protect \BOthers {.}}{%
{\protect \APACyear {1998}}%
}]{%
Stone_1998}
\APACinsertmetastar {%
Stone_1998}%
\begin{APACrefauthors}%
Stone, E\BPBI C.%
, Frandsen, A\BPBI M.%
, Mewaldt, R\BPBI A.%
, Christian, E\BPBI R.%
, Margolies, D.%
, Ormes, J\BPBI F.%
\BCBL {}\ \BBA {} Snow, F.%
\end{APACrefauthors}%
\unskip\
\newblock
\APACrefYearMonthDay{1998}{Jul}{01}.
\newblock
{\BBOQ}\APACrefatitle {The Advanced Composition Explorer} {The advanced
  composition explorer}.{\BBCQ}
\newblock
\APACjournalVolNumPages{Space Science Reviews}{86}{1}{1-22}.
\newblock
\begin{APACrefURL} \url{https://doi.org/10.1023/A:1005082526237}
  \end{APACrefURL}
\newblock
\begin{APACrefDOI} \doi{10.1023/A:1005082526237} \end{APACrefDOI}
\PrintBackRefs{\CurrentBib}

\bibitem [\protect \citeauthoryear {%
{Temmer}%
}{%
{Temmer}%
}{%
{\protect \APACyear {2021}}%
}]{%
Temmer_2021}
\APACinsertmetastar {%
Temmer_2021}%
\begin{APACrefauthors}%
{Temmer}, M.%
\end{APACrefauthors}%
\unskip\
\newblock
\APACrefYearMonthDay{2021}{{\APACmonth{12}}}{}.
\newblock
{\BBOQ}\APACrefatitle {{Space weather: the solar perspective}} {{Space weather:
  the solar perspective}}.{\BBCQ}
\newblock
\APACjournalVolNumPages{Living Reviews in Solar Physics}{18}{1}{4}.
\newblock
\begin{APACrefDOI} \doi{10.1007/s41116-021-00030-3} \end{APACrefDOI}
\PrintBackRefs{\CurrentBib}

\bibitem [\protect \citeauthoryear {%
Thomas%
, Fazakerley%
, Wicks%
\BCBL {}\ \BBA {} Green%
}{%
Thomas%
\ \protect \BOthers {.}}{%
{\protect \APACyear {2018}}%
}]{%
Thomas_2018}
\APACinsertmetastar {%
Thomas_2018}%
\begin{APACrefauthors}%
Thomas, S\BPBI R.%
, Fazakerley, A.%
, Wicks, R\BPBI T.%
\BCBL {}\ \BBA {} Green, L.%
\end{APACrefauthors}%
\unskip\
\newblock
\APACrefYearMonthDay{2018}{}{}.
\newblock
{\BBOQ}\APACrefatitle {Evaluating the Skill of Forecasts of the Near-Earth
  Solar Wind Using a Space Weather Monitor at L5} {Evaluating the skill of
  forecasts of the near-earth solar wind using a space weather monitor at
  l5}.{\BBCQ}
\newblock
\APACjournalVolNumPages{Space Weather}{16}{7}{814-828}.
\newblock
\begin{APACrefURL}
  \url{https://agupubs.onlinelibrary.wiley.com/doi/abs/10.1029/2018SW001821}
  \end{APACrefURL}
\newblock
\begin{APACrefDOI} \doi{https://doi.org/10.1029/2018SW001821} \end{APACrefDOI}
\PrintBackRefs{\CurrentBib}

\bibitem [\protect \citeauthoryear {%
Verscharen%
, Chandran%
, Bourouaine%
\BCBL {}\ \BBA {} Hollweg%
}{%
Verscharen%
\ \protect \BOthers {.}}{%
{\protect \APACyear {2015}}%
}]{%
Verscharen_2015}
\APACinsertmetastar {%
Verscharen_2015}%
\begin{APACrefauthors}%
Verscharen, D.%
, Chandran, B\BPBI D\BPBI G.%
, Bourouaine, S.%
\BCBL {}\ \BBA {} Hollweg, J\BPBI V.%
\end{APACrefauthors}%
\unskip\
\newblock
\APACrefYearMonthDay{2015}{jun}{}.
\newblock
{\BBOQ}\APACrefatitle {DECELERATION OF ALPHA PARTICLES IN THE SOLAR WIND BY
  INSTABILITIES AND THE ROTATIONAL FORCE: IMPLICATIONS FOR HEATING, AZIMUTHAL
  FLOW, AND THE PARKER SPIRAL MAGNETIC FIELD} {Deceleration of alpha particles
  in the solar wind by instabilities and the rotational force: Implications for
  heating, azimuthal flow, and the parker spiral magnetic field}.{\BBCQ}
\newblock
\APACjournalVolNumPages{The Astrophysical Journal}{806}{2}{157}.
\newblock
\begin{APACrefURL} \url{https://dx.doi.org/10.1088/0004-637X/806/2/157}
  \end{APACrefURL}
\newblock
\begin{APACrefDOI} \doi{10.1088/0004-637X/806/2/157} \end{APACrefDOI}
\PrintBackRefs{\CurrentBib}

\bibitem [\protect \citeauthoryear {%
Verscharen%
, Klein%
\BCBL {}\ \BBA {} Maruca%
}{%
Verscharen%
\ \protect \BOthers {.}}{%
{\protect \APACyear {2019}}%
}]{%
Verscharen_2019}
\APACinsertmetastar {%
Verscharen_2019}%
\begin{APACrefauthors}%
Verscharen, D.%
, Klein, K\BPBI G.%
\BCBL {}\ \BBA {} Maruca, B\BPBI A.%
\end{APACrefauthors}%
\unskip\
\newblock
\APACrefYearMonthDay{2019}{Dec}{09}.
\newblock
{\BBOQ}\APACrefatitle {The multi-scale nature of the solar wind} {The
  multi-scale nature of the solar wind}.{\BBCQ}
\newblock
\APACjournalVolNumPages{Living Reviews in Solar Physics}{16}{1}{5}.
\newblock
\begin{APACrefURL} \url{https://doi.org/10.1007/s41116-019-0021-0}
  \end{APACrefURL}
\newblock
\begin{APACrefDOI} \doi{10.1007/s41116-019-0021-0} \end{APACrefDOI}
\PrintBackRefs{\CurrentBib}

\bibitem [\protect \citeauthoryear {%
Verscharen%
\ \BBA {} Marsch%
}{%
Verscharen%
\ \BBA {} Marsch%
}{%
{\protect \APACyear {2011}}%
}]{%
Verscharen_2011}
\APACinsertmetastar {%
Verscharen_2011}%
\begin{APACrefauthors}%
Verscharen, D.%
\BCBT {}\ \BBA {} Marsch, E.%
\end{APACrefauthors}%
\unskip\
\newblock
\APACrefYearMonthDay{2011}{}{}.
\newblock
{\BBOQ}\APACrefatitle {Apparent temperature anisotropies due to wave activity
  in the solar wind} {Apparent temperature anisotropies due to wave activity in
  the solar wind}.{\BBCQ}
\newblock
\APACjournalVolNumPages{Annales Geophysicae}{29}{5}{909--917}.
\newblock
\begin{APACrefURL} \url{https://angeo.copernicus.org/articles/29/909/2011/}
  \end{APACrefURL}
\newblock
\begin{APACrefDOI} \doi{10.5194/angeo-29-909-2011} \end{APACrefDOI}
\PrintBackRefs{\CurrentBib}

\bibitem [\protect \citeauthoryear {%
Wilson%
, Bagenal%
\BCBL {}\ \BBA {} Persoon%
}{%
Wilson%
\ \protect \BOthers {.}}{%
{\protect \APACyear {2017}}%
}]{%
wilson_2017}
\APACinsertmetastar {%
wilson_2017}%
\begin{APACrefauthors}%
Wilson, R\BPBI J.%
, Bagenal, F.%
\BCBL {}\ \BBA {} Persoon, A\BPBI M.%
\end{APACrefauthors}%
\unskip\
\newblock
\APACrefYearMonthDay{2017}{}{}.
\newblock
{\BBOQ}\APACrefatitle {Survey of thermal plasma ions in Saturn's magnetosphere
  utilizing a forward model} {Survey of thermal plasma ions in saturn's
  magnetosphere utilizing a forward model}.{\BBCQ}
\newblock
\APACjournalVolNumPages{Journal of Geophysical Research: Space
  Physics}{122}{7}{7256-7278}.
\newblock
\begin{APACrefURL}
  \url{https://agupubs.onlinelibrary.wiley.com/doi/abs/10.1002/2017JA024117}
  \end{APACrefURL}
\newblock
\begin{APACrefDOI} \doi{https://doi.org/10.1002/2017JA024117} \end{APACrefDOI}
\PrintBackRefs{\CurrentBib}

\bibitem [\protect \citeauthoryear {%
Wilson%
\ \protect \BOthers {.}}{%
Wilson%
\ \protect \BOthers {.}}{%
{\protect \APACyear {2008}}%
}]{%
wilson_2008}
\APACinsertmetastar {%
wilson_2008}%
\begin{APACrefauthors}%
Wilson, R\BPBI J.%
, Tokar, R\BPBI L.%
, Henderson, M\BPBI G.%
, Hill, T\BPBI W.%
, Thomsen, M\BPBI F.%
\BCBL {}\ \BBA {} Pontius~Jr., D\BPBI H.%
\end{APACrefauthors}%
\unskip\
\newblock
\APACrefYearMonthDay{2008}{}{}.
\newblock
{\BBOQ}\APACrefatitle {Cassini plasma spectrometer thermal ion measurements in
  Saturn's inner magnetosphere} {Cassini plasma spectrometer thermal ion
  measurements in saturn's inner magnetosphere}.{\BBCQ}
\newblock
\APACjournalVolNumPages{Journal of Geophysical Research: Space
  Physics}{113}{A12}{}.
\newblock
\begin{APACrefURL}
  \url{https://agupubs.onlinelibrary.wiley.com/doi/abs/10.1029/2008JA013486}
  \end{APACrefURL}
\newblock
\begin{APACrefDOI} \doi{https://doi.org/10.1029/2008JA013486} \end{APACrefDOI}
\PrintBackRefs{\CurrentBib}

\bibitem [\protect \citeauthoryear {%
Wüest%
, Evans%
\BCBL {}\ \BBA {} Steiger%
}{%
Wüest%
\ \protect \BOthers {.}}{%
{\protect \APACyear {2007}}%
}]{%
wuest_2007}
\APACinsertmetastar {%
wuest_2007}%
\begin{APACrefauthors}%
Wüest, M.%
, Evans, D\BPBI S\BPBI D\BPBI S.%
\BCBL {}\ \BBA {} Steiger, R\BPBI v\BPBI R.%
\end{APACrefauthors}%
\unskip\
\newblock
\APACrefYear{2007}.
\newblock
\APACrefbtitle {Calibration of particle instruments in space physics / editors,
  Martin Wüest, David S. Evans, Rudolf von Steiger.} {Calibration of particle
  instruments in space physics / editors, martin wüest, david s. evans, rudolf
  von steiger.}
\newblock
\APACaddressPublisher{Noordwijk, The Netherlands}{Published for the
  International Space Science Institute by ESA Publications Division}.
\PrintBackRefs{\CurrentBib}

\bibitem [\protect \citeauthoryear {%
Yates%
\ \BBA {} Goodman%
}{%
Yates%
\ \BBA {} Goodman%
}{%
{\protect \APACyear {2014}}%
}]{%
yates_2014}
\APACinsertmetastar {%
yates_2014}%
\begin{APACrefauthors}%
Yates, R\BPBI D.%
\BCBT {}\ \BBA {} Goodman, D\BPBI J.%
\end{APACrefauthors}%
\unskip\
\newblock
\APACrefYear{2014}.
\newblock
\APACrefbtitle {Probability and stochastic processes : a friendly introduction
  for electrical and computer engineers / Roy D. Yates, Rutgers, the State
  University of New Jersey, David J. Goodman, New York University.}
  {Probability and stochastic processes : a friendly introduction for
  electrical and computer engineers / roy d. yates, rutgers, the state
  university of new jersey, david j. goodman, new york university.}\
  (\PrintOrdinal{Third edition.}\ \BEd).
\newblock
\APACaddressPublisher{Hoboken, NJ}{John Wiley \& Sons, Inc.}
\PrintBackRefs{\CurrentBib}

\bibitem [\protect \citeauthoryear {%
{Zhang}%
, {Verscharen}%
\BCBL {}\ \BBA {} {Nicolaou}%
}{%
{Zhang}%
\ \protect \BOthers {.}}{%
{\protect \APACyear {2024}}%
}]{%
dataset23}
\APACinsertmetastar {%
dataset23}%
\begin{APACrefauthors}%
{Zhang}, H.%
, {Verscharen}, D.%
\BCBL {}\ \BBA {} {Nicolaou}, G.%
\end{APACrefauthors}%
\unskip\
\newblock
\APACrefYearMonthDay{2024}{{\APACmonth{01}}}{}.
\newblock
\APACrefbtitle {{The impact of non-equilibrium plasma distributions on solar
  wind measurements by Vigil's Plasma Analyser}.} {{The impact of
  non-equilibrium plasma distributions on solar wind measurements by Vigil's
  Plasma Analyser}.}
\newblock
\APACaddressPublisher{}{Zenodo}.
\newblock
\begin{APACrefURL} \url{https://doi.org/10.5281/zenodo.10550337}
  \end{APACrefURL}
\newblock
\begin{APACrefDOI} \doi{10.5281/zenodo.10550337} \end{APACrefDOI}
\PrintBackRefs{\CurrentBib}

\end{thebibliography}
\newpage
\appendix
\section{Effect of $\kappa$-distribution}
\label{appendix_kappa}

A common feature of the solar wind are non-thermal tails in the velocity distribution functions. This distribution is describing plasma out of the classic thermal equillibrium and has been observed in numerous space plasmas \cite{Livadiotis_2009,Livadiotis_2013,Nicolaou_2018}. In this Appendix, we analyze the impact of these tails on the performance of PLA. We represent the relevant input distribution for this analysis as a $\kappa$-distribution: 
\begin{equation}
\label{kappa_equation}
f_\kappa(\vec u)=\frac{N_{\mathrm{in}}}{w^3}\left[\frac{2}{\pi(2 \kappa-3)}\right]^{3/2} \frac{\Gamma(\kappa+1)}{\Gamma\left(\kappa-\frac{1}{2}\right)}\left[1+\frac{2}{2 \kappa-3} \frac{\left(\vec{u}-\vec U_{\mathrm{in}}\right)^2}{w^2}\right]^{-\kappa-1},
\end{equation}
where $\Gamma(x)$ is the $\Gamma$-function, and $\kappa>3/2$ is the $\kappa$-index that describes how far the system is from the classic isotropic thermal equilibrium. The isotropic thermal speed is here defined as $w=\sqrt{2k_BT_{\mathrm{in}}/m_p}$.

\begin{figure}[!h]
\centering
\noindent\includegraphics[width=\textwidth]{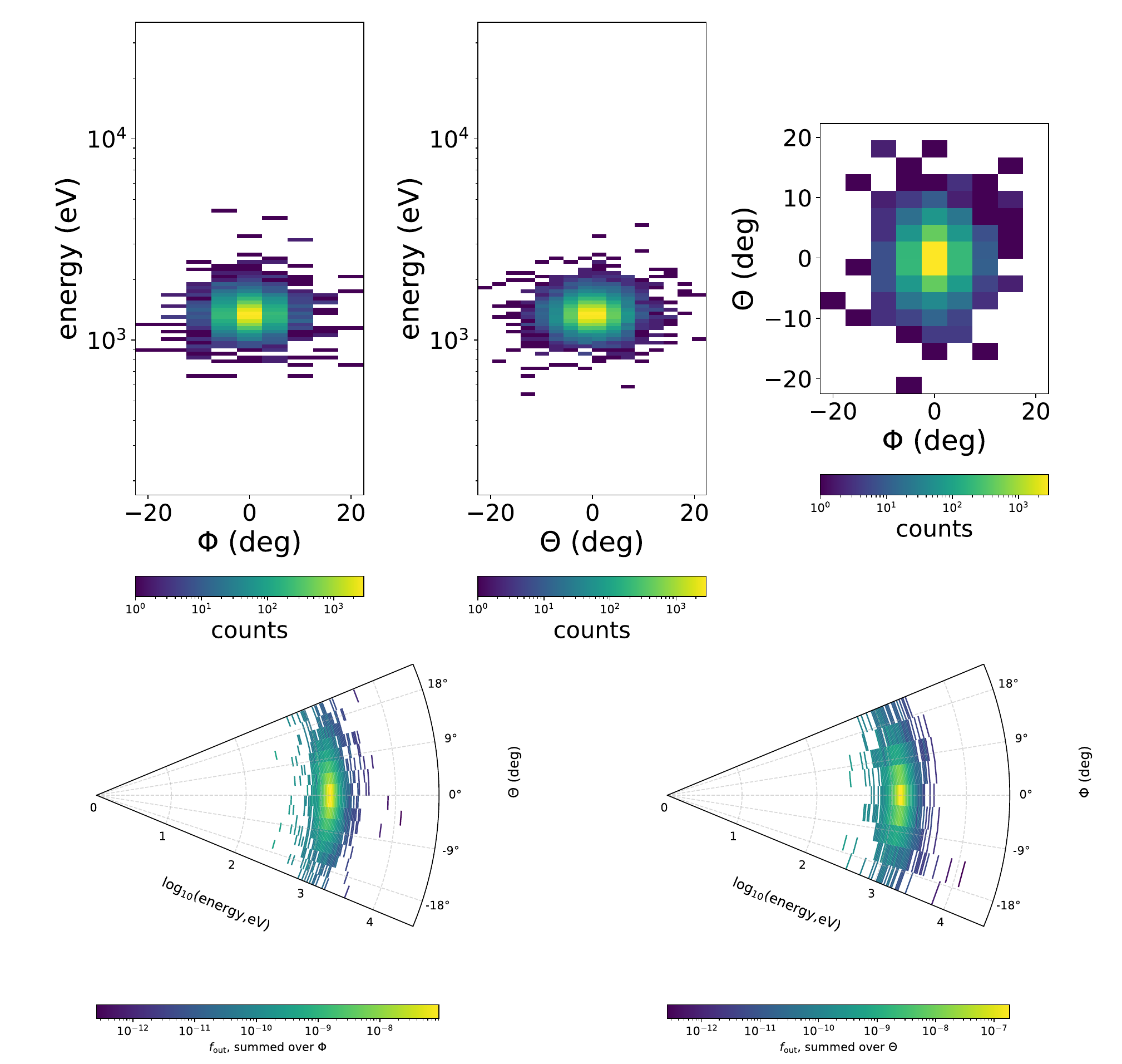}
\caption{
Results of our performance model for an input plasma with a $\kappa$-distribution for $\kappa = 2$ with $N_{\mathrm{in}}=10\,\mathrm{cm}^{-3}$, $U_{x,\mathrm{in}} = 500\,\mathrm{km/s}$, and $T_{\mathrm{in}}=10\,\mathrm{eV}$. Top: Count maps in energy, azimuth, and elevation. Bottom: Output VDF $f_{\mathrm{out}}$ in energy-angle space, summed over the other angle. 
}
\label{k-simulation_2}
\end{figure}

\begin{figure}[!h]
\centering
\noindent\includegraphics[width=\textwidth]{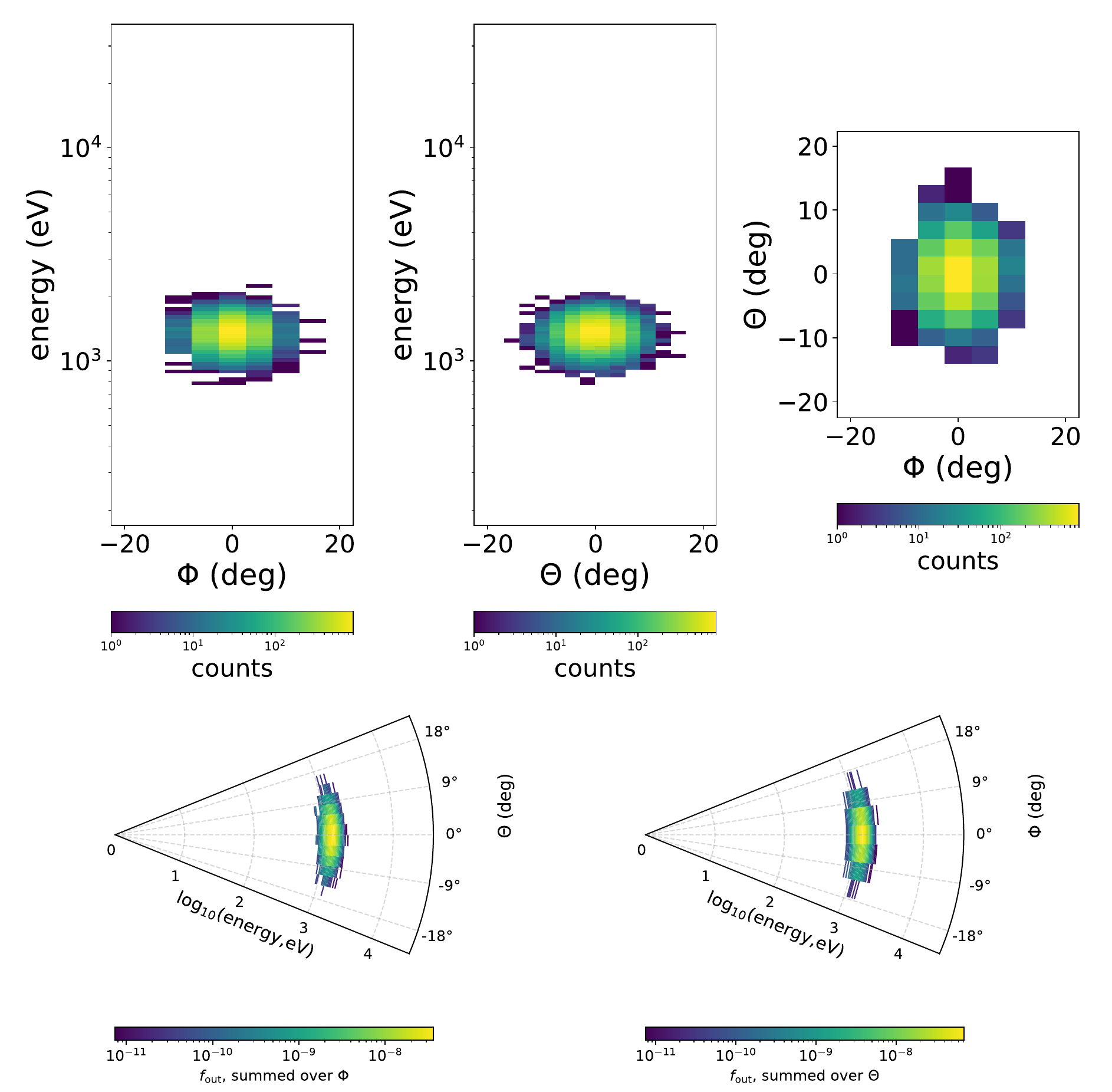}
\caption{
Results of our performance model for an input plasma with a $\kappa$-distribution for $\kappa = 100$ with $N_{\mathrm{in}}=10\,\mathrm{cm}^{-3}$, $U_{x,\mathrm{in}} = 500\,\mathrm{km/s}$, and $T_{\mathrm{in}}=10\,\mathrm{eV}$. Top: Count maps in energy, azimuth, and elevation. Bottom: Output VDF $f_{\mathrm{out}}$ in energy-angle space, summed over the other angle.
}
\label{k-simulation_100}
\end{figure}

Figure~\ref{k-simulation_2} and Figure~\ref{k-simulation_100} show sets of simulated counts and reconstructed VDF maps with $\kappa = 2$ (top) and $\kappa = 100$ (bottom). For a small value of $\kappa$, the particles are dispersed in the field of view and in energy, while for a large $\kappa$, the particles are more concentrated and clustered within the field of view and in energy. We note that the large-$\kappa$ case corresponds to the Maxwellian limit of the $\kappa$-distribution.

\begin{figure}[!h]
\centering
\noindent\includegraphics[width=\textwidth]{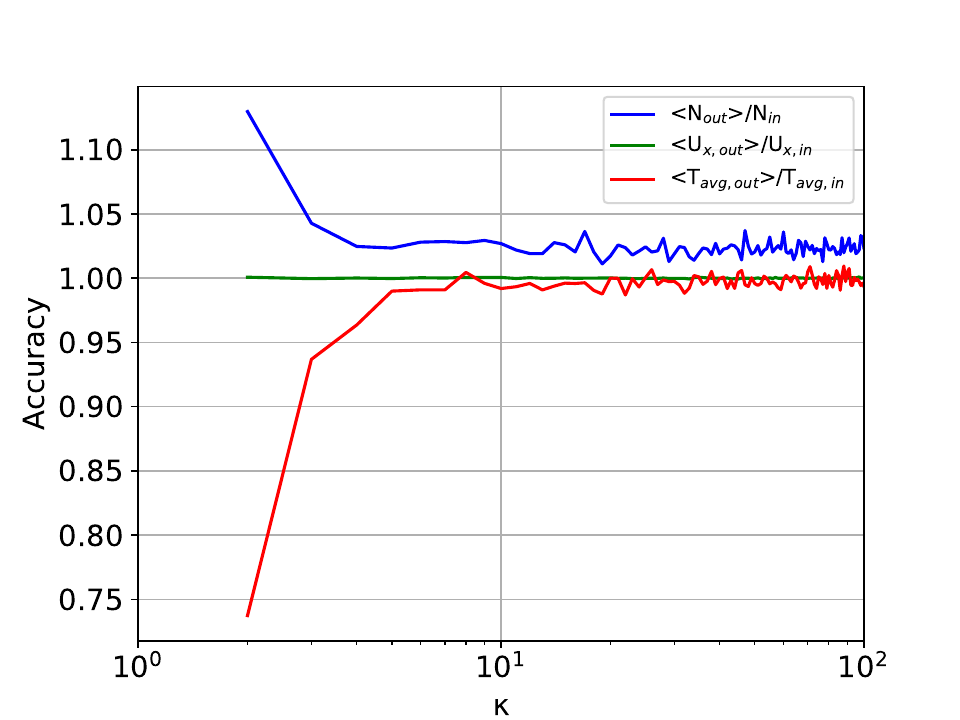}
\caption{Error analysis diagram for a $\kappa$-distributed proton plasma as a function of $\kappa$. The format is the same as in Figure~\ref{bi-max_analysis}.}
\label{k-analysis}
\end{figure}

We show the accuracy curves for a $\kappa$-distributed proton plasma in Figure~\ref{k-analysis}, where we only vary $\kappa$, and the other input moments are fixed to $N_{\mathrm{in}}=10\,\mathrm{cm}^{-3}$, $U_{x,\mathrm{in}} = 500\,\mathrm{km/s}$, and $T_{\mathrm{in}}=10\,\mathrm{eV}$. When $\kappa$ approaches 100, the output moments are very close to the input moments.

In our effort to consider more realistic solar wind distributions, we analyze solar wind protons with a $\kappa$-distribution. We show that, for small $\kappa$, the simulated count maps become more fragmented, causing overestimations of the number density and underestimations of the temperature. For large $\kappa$, the distribution of particles is very close to the Maxwellian distribution \cite<e.g.,>{Livadiotis_2013,Nicolaou_2018,Verscharen_2019}. Our accuracy plot illustrates consistently that the agreement between output and input moments increases as $\kappa$ increases.

\newpage
\section{Effect of $\alpha$-particles}
\label{appendix_alphas}

In this appendix, we discuss the effect of $\alpha$-particles without the addition of a proton beam.

\begin{figure}[!h]
\centering
\noindent\includegraphics[width=\textwidth]{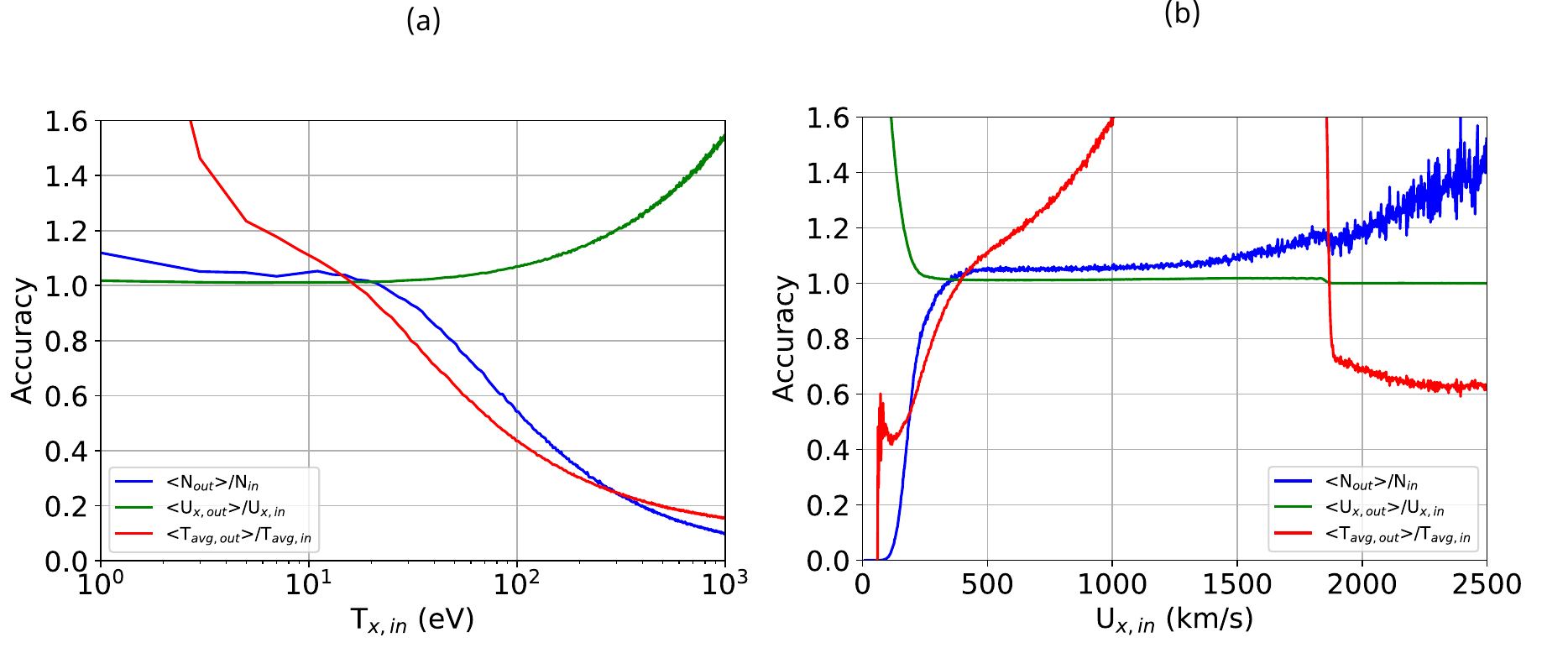}
\caption{
Measurement accuracy for an anisotropic input plasma with a bi-Maxwellian proton core distribution with $N_{\mathrm{in}}=10\,\mathrm{cm}^{-3}$, $U_{y,\mathrm{in}}=U_{z,\mathrm{in}}=0\,\mathrm{km/s}$, and $T_{y,\mathrm{in}}=T_{z,\mathrm{in}}=4T_{x,\mathrm{in}}$, and an added $\alpha$-particle population ($N_{\mathrm{in}}^{\prime\prime}=0.04N_{\mathrm{in}}$, $U_{i,\mathrm{in}}^{\prime\prime}=U_{i,\mathrm{in}}$, and $T_{i,\mathrm{in}}^{\prime\prime}=T_{i,\mathrm{in}}$). Left: Accuracy of  $N_{\mathrm{out}}$ and $U_{x,\mathrm{out}}$ as a function of $T_{x,\mathrm{in}}$ for  $U_{x,\mathrm{in}}=500\,\mathrm{km/s}$. Right: Accuracy of $N_{\mathrm{out}}$ and $U_{x,\mathrm{out}}$ as a function of $U_{x,\mathrm{in}}$ for $T_{x,\mathrm{in}}=10\,\mathrm{eV}$.}
\label{core_alpha_analysis}
\end{figure}

Figure \ref{core_alpha_analysis} shows the accuracy plots under the assumption of an anisotropic proton core and the presence of $\alpha$-particles. The overall trends of the accuracy curves are similar to those of proton temperature anisotropy alone (see section \ref{section:result:Bi-Maxwellian distribution function}). We attribute this similarity to the relatively low number density of input $\alpha$-particles. 

However, we still observe misestimations for cold and slow solar wind. We expect that an increase in the relative number density of $\alpha$-particles would increase this effect. Similar to Section
\ref{section:result:Distribution with alpha particles}, we see a little drop-off for both the number density curve and bulk-velocity curve and a significant drop for temperature curve at high velocities (about 1800\,km/s). This further supports our conjecture (see Section \ref{Discussion:proton beams and alpha particles}) that $\alpha$-particles more easily leave the field of view as their input bulk velocity increases due to their higher energy-per-charge compared to the protons.

\end{document}